\documentclass{article}
\usepackage[a4paper, total={6in, 9in}]{geometry}
\usepackage{stmaryrd}
\usepackage{amsthm,amssymb,amsmath}
\usepackage{color}
\usepackage{array}
\usepackage{url}
\usepackage{listings}
\usepackage{ebproof}
\usepackage{subcaption}
\usepackage{thmtools}
\usepackage{thm-restate}
\usepackage{marvosym}
\usepackage{tcolorbox}
\usepackage{tikz}
\usetikzlibrary{matrix}
\usetikzlibrary{decorations.pathreplacing,calc}
\usepackage{graphicx}
\usepackage[bookmarks,unicode,colorlinks=true,citecolor=violet,urlcolor=blue,linkcolor=red]{hyperref}
\usepackage{float}
\usepackage{paralist}

\newtcbox{\romanobox}[1][gray]{on line,
arc=7pt,colback=#1!35!white,colframe=#1!50!black,
before upper={\rule[-3pt]{0pt}{10pt}},boxrule=0pt,
boxsep=0pt,left=4pt,right=4pt,top=2pt,bottom=2pt}

\newtcbox{\romanotextbox}[1][gray]{on line,
arc=7pt,colback=#1!35!white,colframe=#1!50!black,
before upper={\rule[-3pt]{0pt}{10pt}},boxrule=0pt,
boxsep=0pt,left=1pt,right=1pt,top=2pt,bottom=2pt}

\newtcbox{\romanoframe}[1][gray]{on line,
arc=1pt,colback=white,colframe=black!70,
before upper={\rule[-3pt]{0pt}{10pt}},boxrule=1pt,
}

\newtcbox{\romanoframedeux}[1][gray]{on line,
arc=1pt,colback=white,colframe=black!70,
before upper={\rule[-3pt]{0pt}{10pt}},boxrule=1pt,
boxsep=0pt,left=4pt,right=4pt,top=0pt,bottom=0pt
}

\definecolor{CODEKWD}{RGB}{7,61,111}
\definecolor{CODEFN}{RGB}{14,113,89}
\definecolor{CODECON}{RGB}{7,61,111}
\newcommand{\INDENT}{\quad}

\newcommand{\colorcode}{\colorlet{codekwd}{CODEKWD}\colorlet{codefn}{CODEFN}\colorlet{codecon}{CODECON}}
\colorcode

\newcommand{\kwd}[1]{{\color{codekwd}{\mathtt{#1}}}}

\newenvironment{codeseq}{\begin{array}[t]{@{}>{\tt}l@{}}}{\end{array}}

\makeatletter
\def\ws{}
\long\def\@chang#1{\{\\{\INDENT}\begin{codeseq}#1\end{codeseq}\\\}}
\newcommand{\chang}[1]{#1\ \@ifnextchar\bgroup{\def\ws{}\@chang}{\def\ws{\ \def\ws{}}}}
\let\spkt=;
\let\le=<
\catcode`;=\active
\catcode`<=\active
\def;{\ifmmode\text{\color{codekwd}\texttt{;}}\ {}\else\spkt\fi}
\long\def\@ASGN#1{\mathrel{\kwd{=}}}
\long\def\@SAMPLE#1{\mathrel{\kwd{{*}{=}}}}
\def<{\@ifnextchar -{\@ASGN}{\@ifnextchar *{\@SAMPLE}\le}}
\makeatother

\newcommand{\niceinstr}[1]{\chang{\kwd{#1}}}

\newcommand{\expr}[1]{\mathtt{ #1}}
\newcommand{\e}{\expr{e}}
\renewcommand{\d}{\expr{d}}
\newcommand{\Operator}{\ens{O}}
\newcommand{\Var}{\ens{V}}
\newcommand{\oper}[1]{\mathtt{#1}}
\newcommand{\op}{\oper{op}}
\newcommand{\declass}{\niceinstr{\oper{declass}}\!}
\newcommand{\pad}{\oper{pad}}
\newcommand{\pars}{\oper{parse}}
\newcommand{\sz}{|\cdot|}
\newcommand{\pair}{\oper{cons}}
\newcommand{\et}{\oper{and}}
\newcommand{\ou}{\oper{or}}
\newcommand{\non}{\oper{not}}
\newcommand{\variable}[1]{\mathtt{#1}}

\newcommand{\x}{\variable{x}}
\newcommand{\y}{\variable{y}}
\newcommand{\z}{\variable{z}}

\newcommand{\Command}[1]{\mathtt{ #1}}
\newcommand{\cmd}{\Command{s}}
\newcommand{\prog}{\mathtt{P}}
\newcommand{\X}{\variable{X}}
\newcommand{\Y}{\variable{Y}}
\newcommand{\F}{\variable{F}}

\newcommand{\va}{\variable{a}}

\newcommand{\term}{\variable{t}}
\newcommand{\clos}{\variable{c}}
\newcommand{\proc}{\mathtt{p}}
\newcommand{\gauche}{\mathtt{hd}}
\newcommand{\droite}{\mathtt{tl}}
\newcommand{\truncate}{{\mathtt{truncate}}}
\newcommand{\var}{\texttt{var}\ }
\newcommand{\callcc}{\niceinstr{call}}
\newcommand{\procn}{\mathtt{p}}

\newcommand{\body}{\Command{body}}
\newcommand{\param}{\Command{param}}
\newcommand{\local}{\texttt{local}}

\newcommand{\ens}[1]{\mathbb{#1}}

\newcommand{\instr}[1]{\mathtt{#1}}
\newcommand{\skp}{\niceinstr{skip}}
\newcommand{\sep}{\niceinstr{;}\!\!}
\newcommand{\asg}{\niceinstr{\ :=\ }}
\newcommand{\while}{\niceinstr{while}\!}
\newcommand{\ifor}{\niceinstr{for}\!}
\newcommand{\ito}{\ \niceinstr{to }}

\newcommand{\abs}[1]{{\niceinstr{\lambda}\!#1}}
\newcommand{\ret}{\niceinstr{return}}
\newcommand{\ifa}{\niceinstr{if}\!}

\newcommand{\elsea}{\niceinstr{else}\!}
\newcommand{\brk}{\niceinstr{break}\!}
\newcommand{\main}{\instr{prog}}
\newcommand{\ifar}[3]{\ifa(#1)\{#2\}\elsea\{#3\}}
\newcommand{\whar}[2]{\while(#1)\ \{#2\}}
\newcommand{\brar}[1]{\brk(#1)}
\newcommand{\rgl}{\triangleq}

\newcommand{\A}{\mathcal{A}}

\newcommand{\true}{1}
\newcommand{\false}{0}
\newcommand{\W}{\ens{W}}
\newcommand{\TW}{\texttt{W}}
\newcommand{\STW}{\texttt{T}}
\newcommand{\SN}{\mathrm{TERM}_2}
\newcommand{\TERM}{\mathrm{TERM}}
\newcommand{\ASN}{\dap \cap \SN}

\newcommand{\SST}{\mathcal{T}_\TW}
\newcommand{\w}{\mathit{w}}
\renewcommand{\v}{\mathit{v}}
\renewcommand{\u}{\mathit{u}}
\newcommand{\size}[1]{|#1|}
\newcommand{\sem}[1]{\mbox{$\llbracket #1 \rrbracket$}}
\newcommand{\store}{\mu}

\newcommand{\Imp}{,}
\newcommand{\level}[1]{#1}
\newcommand{\letin}[1]{{\niceinstr{declare}}#1\ \niceinstr{in}}
\newcommand{\levela}{\level{0}}
\newcommand{\levelb}{\level{1}}
\newcommand{\levelc}{\level{2}}

\newcommand{\slat}[1]{\mathbf{#1}}
\newcommand{\sla}{\slat{\tau}}
\newcommand{\slb}{\slat{\eta}}
\newcommand{\slain}{\sla_{in}}
\newcommand{\slaout}{\sla_{out}}
\newcommand{\SL}{\mathbb{N}}
\newcommand{\SLI}{\mathbb{N}_\infty}
\newcommand{\meet}{\sqcap}
\newcommand{\join}{\sqcup}
\newcommand{\ord}{\leq}
\newcommand{\ordst}{<}
\newcommand{\pbl}{\typenv, \typop \vdash^{\slain}_{\slaout}}
\newcommand{\pblp}{\typenv, \typop \vdash^{\slain'}_{\slaout'}}
\newcommand{\pblpp}{\typenv, \typop \vdash^{\slain''}_{\slaout''}}

\newcommand{\pbla}{\typenvs,\typproc,\typop \vdash}

\newcommand{\dord}{\unlhd}

\newcommand{\typenv}{\Gamma}
\newcommand{\typenvs}{\Gamma_\TW}
\newcommand{\typop}{\Delta}
\newcommand{\typproc}{\Omega}
\newcommand{\dom}{\textit{dom}}

\newcommand{\spt}{\mathrm{SPT}}
\newcommand{\mpt}{\mathrm{MPT}}
\newcommand{\opt}{\mathrm{OPT}}

\newcommand{\flr}{\mathrm{FLR}}
\newcommand{\safe}{\mathrm{SAFE}_2}
\newcommand{\saf}{\mathrm{SAFE}}
\newcommand{\forp}{\mathrm{FOR}}
\newcommand{\ap}{\mathrm{AP}}
\newcommand{\dap}{\mathrm{AP}_2}
\newcommand{\FP}{\texttt{FP}}
\newcommand{\Ptime}{{P}}

\newcommand{\FPSPACE}{\texttt{FPSPACE}}
\newcommand{\BFF}{\texttt{BFF}}
\newcommand{\PP}{\texttt{PP}}
\newcommand{\lar}{\leftarrow}
\newcommand{\toe}{\to_\texttt{ex}}
\newcommand{\tos}{\to_\texttt{st}}
\newcommand{\toenv}{\to_\texttt{prg}}
\newcommand{\toexp}{\to_\texttt{exp}}
\newcommand{\order}{\texttt{ord}}
\newcommand{\tost}{\to_\texttt{stm}}
\newcommand{\boite}[2]{{\niceinstr{box}}#1\ \niceinstr{in}#2}

\newcommand{\Span}{\texttt{Span}_{\typenv,\sla}(\store)}
\newcommand{\Spanp}{\texttt{Span}_{\typenv,\sla}(\store')}
\newcommand{\Spanpp}{\texttt{Span}_{\typenv,\sla}(\store'')}
\definecolor{xxx}{rgb}{.7,0,.9}
\tcbset{on line, left=0pt,right=0pt,top=0pt,bottom=0pt,colframe=white, colback=black!20!}

\newif\ifcomments
\commentstrue
\ifcomments
\newcommand{\bruce}[1]{\textcolor{blue}{(Bruce: #1)}}
\else
\newcommand{\bruce}[1]{}
\fi

\begin{document}
\newtheorem{theorem}{Theorem}[section]
\newtheorem{lemma}[theorem]{Lemma}
\newtheorem{proposition}[theorem]{Proposition}
\newtheorem{corollary}[theorem]{Corollary}
\newtheorem{definition}[theorem]{Definition}
\newtheorem{example}[theorem]{Example}
\newtheorem{remark}[theorem]{Remark}
\title{Declassification Policy for Program Complexity Analysis}

\author{Emmanuel Hainry \and Bruce M. Kapron \and Jean-Yves Marion \and Romain P{\'e}choux}
\maketitle

\begin{abstract}
In automated complexity analysis, noninterference-based type systems statically guarantee, via \emph{soundness}, the property that well-typed programs compute functions of a given complexity class, e.g., the class $\FP$ of functions computable in polynomial time. These characterizations are also extensionally \emph{complete} -- they capture all functions -- but are not intensionally complete as some polytime algorithms are rejected. This impact on expressive power is an unavoidable cost of achieving a tractable characterization. 
To overcome this issue, an avenue arising from security applications is to find a relaxation of noninterference based on a \emph{declassification} mechanism that allows critical data to be released in a safe and controlled manner. Following this path, we present a new and intuitive declassification policy preserving $\FP$-soundness and  capturing strictly more programs than existing noninterference-based systems.
We show the versatility of the approach:  it also provides a new characterization of the class $\BFF$ of second-order polynomial time  computable functions in a second-order imperative language, with first-order procedure calls.
Type inference is tractable: it can be done in polynomial time.
\end{abstract}

\renewcommand{\thelstlisting}{\arabic{lstlisting}}

\section{Introduction}
\subsubsection*{Motivations.}
\emph{Noninterference-based type disciplines} have been applied to the field of static analysis to ensure complexity properties of programs (e.g., bounded runtime, bounded memory), noting that a stratification between iterable and not iterable data is key to preventing exponential behavior~\cite{M11}.
The main intuition is as follows: data on which it is safe to iterate cannot increase and are annotated with level $1$ and data that can increase with level $0$. A variable of level $0$ is not safe for iteration, i.e., it  cannot be used to control loops. Data can flow from $1$ to $0$, under the hypothesis that it is passed-by-value. Flows in the opposite direction are strictly prohibited.
Under a termination hypothesis, safety implies polynomial time soundness by ensuring  that the space of level $1$ data is polynomially bounded in the size of the program input.

This approach, also known as \emph{safe recursion}, \emph{ramified recursion},  \emph{cons-free programs}, or \emph{tiering}, depends on the cornerstone works of~\cite{BC92,LeivantMar93} that provided the first implicit characterizations of  (first-order) polynomial time $\FP$.
The methodology was also adapted not only to characterize polytime~\cite{J01,M11,BKS23,CS14} but also other well-known complexity classes, such as (first-order) polynomial space $\FPSPACE$~\cite{J01,HMP13}, second-order polynomial time $\BFF$~\cite{HKMP20}, or probabilistic polynomial time $\PP$~\cite{DLKO21}.  
The main advantage of the underlying type systems is the tractability of their type inference, which can be done in polynomial time. Consequently, it has led to practice-oriented extensions such as the development of type systems for Java programs to statically ensure that their heap and their stack have a polynomially bounded size or that their runtime is polynomially bounded~\cite{HP18,HP23}. The software {\sc ComplexityParser} introduced in~\cite{HJPZ21} ensures this kind of property.

This main advantage also implies their main weakness: while these characterizations are \emph{sound} and \emph{ extensionally complete}, i.e., all functions of the target complexity class are captured, they are not \emph{ intensionally complete}. This means that there exist false negatives, e.g., algorithms that run in polynomial time but are wrongly rejected. This incompleteness is no great surprise for the theoretician who knows, for example, that the set of programs running in polynomial time is not decidable (and in fact has been proven to be $\Sigma^0_2$-complete in~\cite{H79}) and that a tractable characterization of FP must therefore reject some of these polytime programs.

Since then, a number of studies have sought to increase the expressive power of these characterizations. For example,~\cite{HP23} extended  levels to any positive integer and has program flows to pass-by-reference data by combining noninterference-based typing with shape-analysis.
Despite the extensions made to date, the restrictions encountered with the complexity framework can be too rigid, similar to the problems encountered with traditional information-flow security and privacy applications of noninterference~\cite{VIS96,BP03}. The security community has tried to solve this issue by introducing relaxed versions, as in~\cite{Myers99,VolpanoS00} which, in order to provide more expressive power allow some forms of information flow that violate the noninterference principle, by making it possible to release some information in a controlled way.
Such control takes the form of \emph{declassification} or downgrading policies allowing for instance the publishing of  some part of the secret data. A classic example of declassification is hashing: a password is strictly private but its hash must be shared for password verification to work. Allowing intentional information release while avoiding attack vectors is a challenge that has been studied under various frameworks.
In this respect, \cite{SabelfeldS09} presents a list of principles that declassification should preserve and a taxonomy of those frameworks on \emph{what} can be declassified, \emph{who} may receive the declassified data, \emph{where} declassification is allowed, and \emph{when}, in the sense of how often, or under which complexity constraints.

A natural question, then, is whether similar declassification policies can be defined and implemented in the context of complexity analysis.
At present, the only declassification options under consideration are far too limited.~\cite{HP18} considered declassification outside loops. Hence only a constant number of declassifications can be performed at runtime. 
In \cite{M11}, declassification is \emph{a priori} restricted to a given type level, \emph{à la}~\cite{MyersL00}.
In general, declassifying any predetermined finite number of times will be acceptable as it preserves polynomial soundness; but declassifying data in a loop can lead to exponential program behavior. Indeed, declassifying one single bit of data is not a big deal but repeating this process breaks polytime soundness.

\addtolength{\textfloatsep}{-0.5cm}
\begin{figure}[t]
\begin{minipage}[t]{0.42\textwidth}
\begin{lstlisting}[caption={exponential 1}, captionpos=b, label={ex:exp1},numbers=none]
//x,y,z: unary
y$^0\asg$1$^0\sep$
$\while$(x$^1$>0){
  z$^1\asg\declass$(y$^0$)$^1\sep$
  $\while$(z$^1$>0){
      y$^0\asg$y+1$^0\sep$
      z$^1\asg$z-1$^1$
  }$\sep$
  x$^1\asg$x-1$^1$
}
\end{lstlisting}
\end{minipage}
	\begin{minipage}{0.12\textwidth}
		\phantom{X}
	\end{minipage}
\begin{minipage}[t]{0.42\textwidth}
\begin{lstlisting}[caption={exponential 2}, captionpos=b, label={ex:exp2},numbers=none]
//y: binary, x: bool
x$^1\asg$true$^1\sep$
$\while$(x$^1$){
  y$^0\asg$y-1$^0\sep$
  x$^1\asg\declass$(y$^0$>0)$^1$
} 
\end{lstlisting}
\begin{align*}
\quad &\text{level 0: increasing/noniterable}\\
&\text{level 1: nonincreasing/iterable}
\end{align*}
\end{minipage}
\end{figure}

To illustrate the danger linked with declassification when trying to control polynomial complexity, let us introduce two code samples in Listings~\ref{ex:exp1} and~\ref{ex:exp2}, where the syntactic elements are annotated by their level as a superscript. Declassification is used as a primitive construct to raise data from $0$ to $1$.
The program of Listing~\ref{ex:exp1} typechecks wrt. a control-flow based policy with 2 levels but computes the exponential $y=2^x$ in unary, hence in exponential time. This illustrates why the addition of declassification should not be treated superficially at the risk of breaking polytime soundness.
Naively, we might think that it is enough to limit information leaks to a restricted number of bits to avoid such examples. This hope is dashed by the example of Listing~\ref{ex:exp2}, one single loop manipulating a binary number \texttt{y} where the declassification only leaks a boolean value:
this loop explores all integers smaller than y, hence is exponential in the size of y.

\addtolength{\textfloatsep}{+0.5cm}

There is therefore a need not only to develop declassification techniques that preserve the soundness of noninterference policies for complexity analysis, but also to demonstrate that they can be used to drastically increase the expressive power of these methods.

\subsubsection*{Contributions.}
We introduce a new declassification policy for complexity analysis.
This policy preserves polynomial time soundness and answers a question already posed in~\cite{M11} for having a more general form of declassification.
This policy consists of two ingredients; a \texttt{declass} operator and an \emph{aperiodicity} condition.
Similarly to~\cite{Myers99}, \texttt{declass} takes 2 arguments: the expression to declassify and an expression whose level will bound the level up to which the value will be declassified.
In effect, the declassified value loses some information as the output is a unary representation of the length of the first argument and is bounded in size by the length of the second argument.
The aperiodicity condition checks that loops do not encounter a configuration more than once, hence do not enter a periodic state.
This is necessary to prevent abuses such as in Listing~\ref{ex:exp2} above.
We can say that the declassification policy acts on the \emph{where} dimension in the taxonomy of~\cite{SabelfeldS09}.
We illustrate the use of this declassifying policy and its tamed power in two programming languages: 
\begin{inparaenum}[i)]
\item a simple imperative language with a while loop construct, 
\item a programming language with second-order imperative procedures.
\end{inparaenum}
The declassification policy is expressed through a type system, which defines a \emph{safety} property. 
A safe program is a program that can be typed.
Additional properties are required to ensure polynomial time soundness: \emph{aperiodicity} and, in the second-order case, \emph{termination}. A program is aperiodic if in the guard of each while loop, no two equivalent states can be reached in (the depth of) the derivation tree; a program terminates if it halts on all inputs. Aperiodicity  implies termination on safe first-order programs and this explains why termination is only required for second-order programs. 

The main contributions of this paper are:
\begin{itemize}
\item a new declassification policy (Figure~\ref{fig:t1ts}), called safety,  for program complexity analysis, that corresponds to a partial release of information: only the length of data can be declassified. This policy extends the expressive power of previous work~\cite{M11} by allowing to consider programs with restricted interference.
\item a new characterization of the class of polytime computable functions $\FP$ that captures strictly more programs (like Example~\ref{ex:bubble}) than the ones implementing a strict noninterference policy. All functions in $\FP$ can be expressed by safe and aperiodic programs (Theorem~\ref{thm:fpcomplete}) and, conversely, only polynomial time computable functions can be expressed this way (Theorem~\ref{thm:poli}).
\item a proof that safety is \emph{tractable}: it can be decided in polynomial time (Theorem~\ref{thm:tiproc}).
\item a proof that aperiodicity is $\Pi^0_1$-complete (Theorem~\ref{thm:ap}). Hence, aperiodicity is not more difficult than termination, known to be $\Pi^0_2$-complete~\cite{EGSZ11}, used as a condition in previous work~\cite{M11}. Moreover, we show in Theorem~\ref{thm:for} that a decidable and completeness-preserving criterion trivially ensuring aperiodicity can be designed at the price of reduced expressive power. Consequently, our methodology is automatable.
\item an application of the declassification policy: a new characterization of the class of second-order polytime computable functions $\BFF$ implementing the \emph{Strongly PolyTime} ($\spt$) criterion of~\cite{KS18}. In this setting, safety remains tractable (Theorem~\ref{thm:tiprog}) and aperiodicity is $\Pi^0_1$-hard (Corollary~\ref{coro:ap}).
\end{itemize}

\subsubsection*{Application example of the declassification policy.}
One of the latest characterizations of $\BFF$, introduced in~\cite{KS18}, cannot be obtained without declassification. $\spt$ is defined as the class of second-order functionals computable by an oracle Turing machine with 
\begin{inparaenum}[i)]
\item \emph{Polynomial Step Count}: the machine works in time polynomial in the size of the input and the maximum size of an oracle answer, 
\item \emph{Finite Length Revision}: there exists a natural number $n$ s.t. for any oracle and any input, in the run of the machine, it happens at most $n$ times that an oracle answer has size that exceeds the size of all previous answers. 
\end{inparaenum}
The class of functionals  with polynomial step counts is named $\opt$ (for \emph{Oracle Polynomial Time}, see~\cite{C92}), and the class of functionals with finite length revision is named $\flr$. By definition, $\spt = \opt \cap \flr$. Let $\lambda(X)_2$ be the restriction to second-order functions of the simply-typed lambda-closure of terms with constants in the set $X$. The class $\BFF$ can be recovered from $\spt$ as follows:
\begin{theorem}[\cite{KS18}]\label{thm:KS}
$\lambda(\spt)_2 =  \BFF$.
\end{theorem}

As the above characterization relies on a purely semantic criterion ($\flr$) and only deals with machines, it was an open issue to know to what extent a similar and tractable characterization of $\BFF$ could be obtained for a programming language using noninterference-based analysis.  However this issue cannot be resolved in a strictly noninterfering system for the simple reason that $\spt$ needs interferences. Indeed, in the definition of $\flr$ the control flow needs to be guarded by the oracle answers (or at least a counter that keeps information on the number $n$ of times an answer size has exceeded the size of all previous answer). Hence the loop can be controlled by increasing data (oracle answers are not known in advance and cannot be trusted), which breaks noninterference.
In other words, since in $\spt$, it is the number of increases for the outputs of oracles that needs to be observed,
those outputs need to be declassified in order to manage the control flow.
Therefore, an application of interest is whether declassification techniques could help in capturing the 
$\spt$ characterization. We give a positive answer to this question in the paper.

\subsubsection*{Related work.}
Our results build upon work on declassification, noninterference type-based approaches in implicit computational complexity, and $\BFF$ characterizations.
The typing system we define has roots in~\cite{VIS96} and the declassification policy was inspired from works on downgrading, endorsement, and declassification~\cite{VolpanoS00,LiZ05,
SabelfeldS09,ZdancewicM01}.
In contrast, our aim is to characterize tractable programs rather than prove security or confidentiality properties, which accounts for the additional restrictions of aperiodicity and termination.
On the other side, our framework provides the possibility of declassifying in a loop which is challenging when only partial release of data is allowed.
The implicit complexity line of work draws from  noninterference ideas applied  to  data ramification/safe recursion~\cite{BC92,LeivantMar93} to characterize complexity classes such as $\FP$ or $\FPSPACE$, e.g.,~\cite{M11, J01}.

In contrast to~\cite{C92,KC91,CU93,KS18}, that study $\BFF$ on machines, the current work characterizes $\BFF$ in the setting of programming languages. Hence it follows the approach of~\cite{IRK01,DR06} but differs from the latter in the sense that the bounds do not need to be explicitly provided in the language. This work is also linked to higher-order complexity analysis based on costs and potentials of~\cite{DPR13}, although the theoretical and practical objectives differ.
\cite{HKMP20,HKMP22} provide characterizations of $\BFF$ based on another noninterfering criterion (called $\mpt$ in~\cite{KS18}) and, hence, do not capture algorithmic schemes based on $\spt$. Several work~\cite{HP20,BDKV24} have designed alternative elegant characterizations of $\BFF$ using higher-order polynomial interpretations. However, contrarily to safety, these interpretation methods are not tractable.

\begin{figure}[t]
\hrulefill
\[
\begin{array}{rrcl}
\texttt{Expr} & \e   & := &  \x\ |\ \op(\bar{\e})\ |\ \declass(\e, \e) \\
\texttt{Stmt} & \cmd & := & \skp\ |\ \x \asg \e\ |\ \cmd \sep \cmd\ \ | \  \ifar{\e}{\cmd}{\cmd}\ |\ \\
& & & \whar{\e}{\cmd}\ |\ \brar{\e} \\
\texttt{Prg} & \prog & := & \main(\bar{\x})\{\cmd\ \ret \x\} 
\end{array}\]
\hrulefill
\caption{Syntax}
\label{fig:t1syn}
\end{figure}

\section{Declassifying policy for complexity}\label{s:declass}
In this section, we show how declassification for complexity analysis can be adapted to a simple imperative programming language.

\subsection{Imperative Language with Declassification}\label{sec:t1}
\subsubsection*{Syntax.} 
The syntax of the considered imperative language is provided  in Figure~\ref{fig:t1syn}, where $\x$ is a variable in a fixed set of variables $\Var$ and where $\bar{\e}$ (resp: $\bar{\x}$) represents a finite sequence of expressions (resp: variables). Let $\ell(\bar{\x})$ be the length of the sequence $\bar{\x}$.

Operators $\op$ are basic operations of fixed arity $ar(\op)$ in the set $\Operator$. We assume that $\Operator$ includes some basic boolean and arithmetic operators such as  $\{=, <, \leq, 0, +1, -1, \non, \et, \ou\}$. Any constant can be viewed as an operator of arity $0$. 

An \emph{expression} $\e$ can be a variable $\x$, an operator application $\op(\bar{\e})$ or the result of a declassification $\declass(\e_1, \e_2)$. Operators will be used in infix, postfix, or prefix notations and they will always be fully applied, i.e., in $\op(\bar{\e})$, it always holds that $\ell(\bar{\e})=ar(\op)$.
We will say that an expression is \emph{declassified}, if it appears as a subexpression of some $\e_1$ in $\declass(\e_1,\e_2)$. 

Statements in \emph{Stmt} are standard imperative constructs with assignments, conditionals, loops, and breaks.

Finally, a \emph{program} takes a list of parameter variables $\bar{\x}$, executes a (body) statement $\body(\prog) \triangleq \cmd$, and returns one output variable $\x$.

\begin{example}\label{ex:t1}
In Listing~\ref{bubble}, we have implemented bubble sort of a string with some type annotations as superscript. Bubble uses the following arity-2 operators: \verb:+: to append a character to a string, \verb:<: to compare two characters, \verb:hd: and \verb:tl: to get respectively the first character and the rest of the string. An explicit call to $\declass$ is performed on line $14$ inside a loop and we will see shortly (see Example~\ref{ex:bubble}) that the program cannot type without this feature. Such a program cannot be typed through the discipline of \cite{M11,HP18}. 
\end{example}

\begin{figure*}[t]
	\begin{minipage}{\textwidth}
\hrulefill
\[
\begin{prooftree}
\hypo{\phantom{(\store, \overline{\e}) \toe \overline{\w}}}
\infer1{(\store, \x) \toe \store(\x)}
\end{prooftree}
\qquad
\begin{prooftree}
\hypo{(\store, \overline{\e}) \toe \overline{\w}}
\infer1{(\store, \op(\overline{\e})) \toe \sem{\op}(\overline{\w})}
\end{prooftree}
\qquad 
\begin{prooftree}
\hypo{(\store, \e_1) \toe \w_1}
\hypo{(\store, \e_2) \toe \w_2}
\infer2{(\store, \declass(\e_1, \e_2)) \toe 1^{\min(|\w_1|, |\w_2|)}}
\end{prooftree}
\]

\[
\begin{prooftree}
\hypo{\phantom{(\store, \e_i) \toe \w_1}}
\infer1{(\store, \skp) \tos (\top, \store)}
\end{prooftree}
\qquad
\begin{prooftree}
\hypo{(\store, \e) \toe \w}
\infer1{(\store, \x \asg \e)\tos (\top, \store[\x \lar \w])}
\end{prooftree}
\]

\[
\begin{prooftree}
\hypo{(\store, \cmd_1)\tos(\top, \store')}
\hypo{(\store', \cmd_2)\tos(\square, \store'')}
\infer2[($\square \in \{\top,\bot\}$)]{(\store, \cmd_1 \sep \cmd_2) \tos (\square, \store'')}
\end{prooftree}
\qquad
\begin{prooftree}
\hypo{(\store, \cmd_1)\tos(\bot, \store')}
\infer1{(\store, \cmd_1 \sep \cmd_2) \tos (\bot, \store')}
\end{prooftree}
\]

\[
\begin{prooftree}
\hypo{(\store, \e) \toe \w}
\hypo{(\store, \cmd_{\w}) \tos (\square, \store')}
\infer2[($\square \in \{\top,\bot\}, \w \in \{\underline{\false}, \true\}$)]{\ifar{\e}{\cmd_{\true}}{\cmd_{\underline{\false}}} \tos (\square, \store')}
\end{prooftree}
\qquad
\begin{prooftree}
	\hypo{(\store, \e) \toe \underline{\false}}
\infer1{(\store, \whar{\e}{\cmd}) \tos (\top, \store)}
\end{prooftree}
\]

\[
\begin{prooftree}
\hypo{(\store, \e) \toe \true}
\hypo{(\store, \cmd \sep \whar{\e}{\cmd}) \tos (\square, \store')}
\infer2[($\square \in \{\top,\bot\}$)]{(\store, \whar{\e}{\cmd}) \tos (\top, \store')}
\end{prooftree}
\]

\[
\begin{prooftree}
	\hypo{(\store, \e) \toe \underline{\false}}
\infer1{(\store, \brar{\e}) \tos (\top, \store)}
\end{prooftree}
\qquad
\begin{prooftree}
\hypo{(\store, \e) \toe \true}
\infer1{(\store, \brar{\e}) \tos (\bot, \store)}
\end{prooftree}
\]
\hrulefill
	\end{minipage}
\caption{Semantics}
\label{fig:t1sem}
\end{figure*}

\begin{lstlisting}[basicstyle=\small \ttfamily,linewidth=0.92\textwidth, caption={ Bubble sort}, label={bubble}, captionpos=b,float=h,belowcaptionskip=-21pt]
bubble(list$^1$){
  list1$^\levelb\asg$list$^1\sep$
  len$^0\asg$0$^0\sep$
  $\while$(list1$^1$ $\neq$ $\epsilon$){
    list1$^1\asg$tl(list1)$^1\sep$
    len$^0\asg$(len+1)$^0$
  }$\sep$
  list2$^0\asg$list$^1\sep$
  len1$^1$$\asg$len$^0\sep$
  $\while$(len1$^1$ > 0){
    r$^0\asg\varepsilon^0$$\sep$
    x$^0\asg$hd(list2)$^0\sep$
    list2$^0\asg$tl(list2)$^0\sep$
    len2$^1\asg\declass$(len$^0$,list$^1$)$\sep$
    $\while$(len2$^1$>1){
      y$^0\asg$hd(list2)$^0\sep$
      list2$^0\asg$tl(list2)$^0\sep$
      $\ifa$(x < y)$^0${
        r$^0\asg$(r+x)$^0\sep$
        x$^0\asg$y$^0$
      }$\elsea\!\!${
        r$^0\asg$(r+y)$^0$
      }$\sep$
      len2$^0\asg$(len2-1)$^0$
    }
    r$^0\asg$(r+x)$^0\sep$
    list2$^0\asg$r$^0\sep$
    len1$^1\asg$(len1-1)$^1$
  }
  $\ret$r$^0$
}
\end{lstlisting}

\subsubsection*{Semantics.} Let $\W\triangleq \Sigma^{*}$ be the set of words, also called \emph{values}, over the finite alphabet $\Sigma$ such that $\{0, 1\} \subseteq \Sigma$. 
$\epsilon$ denotes the empty word; $\v\cdot\w$ denotes the concatenation of words $\v$ and $\w$; for $n \in \mathbb{N}$, $\v^n$ is defined inductively as $\v^0\triangleq \epsilon$ and $\v^{n+1}\triangleq \v\cdot\v^{n}$.
Let $\dord$ be the sub-word relation defined by $\v\dord \w$ if there exist words $\u$ and $\u'$ such that $\w=\u\cdot\v\cdot\u'$.
The size of word $\w$ is denoted by $|\w|$.

The semantics assigns a total function $\sem{op}: \W^{ar(\op)}\to \W$  to each operator $\op\in\Operator$.
For example, $+1$ and $-1$ are operators on unary numbers, computing increment $\sem{+1} : \W \to \W$ and decrement $\sem{-1}:\W \to \W$ , respectively,  and outputting $\epsilon$ in case of failure (if the input word is not a unary number). I.e., for $n \in \mathbb{N}$, $\sem{+1}(1^n)\triangleq 1^{n+1}$ and $\sem{+1}(\w)\triangleq \epsilon$, if $\w \neq 1^n$.
The output of Boolean operators is either false ($0$) or true ($1$). 
By extension, any value that is not $\true$, including $\epsilon$, will be denoted by $\underline{\false}$ and considered as false.
The semantics assigned to comparison operators is the standard shortlex order.

A \emph{memory store} $\store\in\texttt{Stores}$ is a total map from variables in $\Var$ to values in $\W$.
Let $dom(\store)$ be the domain of the store $\store$.
Given $\x \in \Var$, and a value $\w\in\W$, let $\store[\x \leftarrow \w]$ denote the store obtained from $\store$ by updating the value of $\x$ to $\w$.
This notation is extended naturally to sequences $\store[\bar{\x} \leftarrow \bar{\w}]$, whenever $\ell(\bar{\x})=\ell(\bar{\w})$.
Finally, let $\store_\emptyset$ denote the empty store, mapping each variable to $\epsilon$.

The operational semantics is described in Figure~\ref{fig:t1sem} as a standard big-step semantics with control flow breaks.
The semantics of expressions and statements are defined by the maps 
\begin{align*}
 \toe &: (\texttt{Stores}\times\texttt{Expr}) \to \W \\
 \tos &: (\texttt{Stores} \times \texttt{Stmt})\to (\{\top, \bot\}\times\texttt{Stores}).
\end{align*}
Evaluating a $\declass(\e_1,\e_2)$ yields the value $1^{\min(|\w_{1}\!|, |\w_2\!|)}$, provided that $\e_i$ evaluates to the value $\w_i$, for $i \in \{1,2\}$.
Though this evaluation treats both operands symmetrically, as we will see shortly, the first operand $\e_1$ is the one to declassify and the second operand $\e_2$ bounds the length that is transmitted.
When evaluating a guard expression $\e$, assumed to evaluate to a boolean, we write $(\store, \e) \toe \underline{\false}$ to denote that $(\store, \e) \toe \w$ with $\w \neq \true$.

The first component in the codomain of $\tos$ is a flag encoding whether the flow has to be broken ($\bot$) or not ($\top$). 
In rules describing the semantics of a sequence $\cmd_1\sep \cmd_2$, it is implicitly assumed that $\cmd_1 \neq \cmd'_1 \sep \cmd''_1$.

For a \emph{configuration} $c \in ({\tt Stores} \times {\tt Stmt}) \cup (\texttt{Stores}\times\texttt{Expr})$, let $\pi_c$ be the \emph{(evaluation) tree with root} $c$ obtained by evaluating $c$ using the rules of Figure~\ref{fig:t1sem}. $\pi_c$ can be infinite if the program does not terminate. 
We write $\pi_c \sqsubseteq \pi_{c'}$ (resp. $\pi_c \sqsubset \pi_{c'}$) when $\pi_c$ is a (strict) subtree of $\pi_{c'}$.
\begin{figure*}[!t]
	\begin{minipage}{\textwidth}
\hrulefill
\[
\begin{prooftree}
\hypo{\typenv(\x)=\sla}
\infer1[(VAR)]{\pbl \x: \sla}
\end{prooftree}
\qquad
\begin{prooftree}
\hypo{\sla_1 \to \cdots \to \sla_{ar(\op)+1} \in \Delta(\op)(\slain, \slaout)}
\hypo{\forall i \leq ar(\op),\ \pbl \e_i: \sla_i}
\infer2[(OP)]{\pbl \op(\bar{\e}): \sla_{ar(\op)+1} }
\end{prooftree}
\]

\[
\begin{prooftree}
\hypo{\pbl \e_1: \sla_1}
\hypo{\pbl \e_2: \slaout}
\hypo{\sla_1 \ord \sla \ord \slaout}
\infer3[(DCL)]{\pbl  \declass(\e_1,\e_2): \sla}
\end{prooftree}
\qquad
\begin{prooftree}
\hypo{\pbl  \cmd\ : \sla_1}
\hypo{\sla_1 \leq \sla_2}
\infer2[(SUB)]{\pbl  \cmd\ : \sla_2}
\end{prooftree}
\]

\[
\begin{prooftree}
\hypo{\phantom{\pbl  \cmd\ : \slb}}
\infer1[(SKP)]{\pbl \skp\ : \levela}
\end{prooftree}
\qquad
\begin{prooftree}
\hypo{\typenv(\x)= \sla_1}
\hypo{\pbl \e: \sla_2}
\hypo{(\slaout = \levela) \vee (\sla_1  \ord \sla_2)}
\infer3[(ASG)]{\pbl \x \asg \e \ : \sla_1}
\end{prooftree}
\]

\[
\begin{prooftree}
\hypo{\pbl  \cmd_1 :\sla}
\hypo{\pbl \cmd_2:\sla}
\infer2[(SEQ)]{\pbl \cmd_1 \sep \cmd_2 \ :\sla}
\end{prooftree}
\]

\[
\begin{prooftree}
\hypo{\pbl \e: \sla}
\hypo{\pbl \cmd_{\true}: \sla}
\hypo{\pbl \cmd_{\false}:\sla}
\infer3[(CND)]{\pbl \ifa (\e) \{\cmd_{\true}\} \elsea \{\cmd_{\false}\}\ :\sla}
\end{prooftree}
\]

\[
\begin{prooftree}
\hypo{\typenv,\typop \vdash^\sla_{\slaout} \e: \sla}
\hypo{\typenv,\typop \vdash^\sla_{\slaout} \cmd : \sla}
\hypo{\levelb \ord \sla \ord \slaout}
\infer3[(WH)]{\pbl  \while (\e) \{\cmd\}\ :\sla}
\end{prooftree}
\]

\[
\begin{prooftree}
\hypo{\typenv,\typop \vdash^{\sla}_\sla \e: \sla}
\hypo{\typenv,\typop \vdash^\sla_\sla \cmd : \sla}
\hypo{\levelb \ord \sla}
\infer3[(WI)]{\typenv,\typop \vdash^{\levela}_\levela  \while (\e) \{\cmd\}\ :\sla}
\end{prooftree}
\qquad
\begin{prooftree}
\hypo{\pbl \e  : \sla}
\hypo{ \slain \ord \sla}
\infer2[(BRK)]{\pbl \brk(\e)  : \slain}
\end{prooftree}
\]
\hrulefill
	\end{minipage}
\caption{Level-based typing rules}
\label{fig:t1ts}
\end{figure*}

\subsubsection*{Restriction on Operators.}
Following~\cite{M11}, we define three classes of operators called neutral, positive, and polynomial depending on the total function they compute. 
This classification of operators will be used by the type system as the admissible types for operators will depend on their category.

\begin{definition}
\label{def:np}
An operator $\op \in \Operator$ of arity $k$ is:\\
--\emph{neutral} if:
1) either $\sem{\op} \in \W^{k} \to \{\false, \true\}$,\\
\phantom{--\emph{neutral} if:} 2) or  $\exists i  \leq k,\ \forall \bar{\w} \in \W^{k},\ \sem{\op}(\bar{\w}) \dord {\w_i}$;
\\
-- \emph{positive} if $\exists c_{\op} \in \mathbb{N}, \ \forall \bar{\w} \in \W^{k},\ \size{\sem{\op}(\bar{\w})}  \leq \max_i \size{\w_i} + c_{\op}
$;\\
--\emph{polynomial} if $ \exists P \in \mathbb{N}[X],\ 
\forall \bar{\w} \in \W^{k},\ \size{\sem{\op}(\bar{\w})}  \leq P(\max_i \size{\w_i})
$.
\end{definition}

The basic intuition behind this classification is that neutral operators are iterable; positive operators are polynomially iterable; and polynomial operators are not iterable. Each operator that does not fall in these categories will be rejected by the type system.
A neutral operator is always positive and a positive operator is always polynomial but the converse properties are not true.
In the sequel, we reserve the name positive (resp. polynomial) for those operators that are positive but not neutral (resp. polynomial but not positive). 

\begin{example}\label{ex:npp}
In Listing~\ref{bubble}, operators $\neq$, $>$, $<$ compute boolean predicates, hence they are \emph{neutral}.
\verb:hd:, \verb:tl:, and $-1$ (unary) are subword operations, as such, they are neutral.
$+1$ adds one character, it is hence a positive operator (with $c_{+1}$ equal to $1$).
The $+$ operator (line 26) takes a word $\w \in \W$ and a character $a \in \Sigma$ and outputs the word $\w\cdot a$, it is hence a positive operator as $\size{\w\cdot a}=\size{\w}+1$.
\end{example}

\subsection{Type System and Safe Programs}\label{ss:type}
We now describe the type discipline implementing the noninterference policy with declassification.
\subsubsection*{Levels and Typing Environments.}
\emph{Levels} are types that belong to the set of non-negative integers $ \SL$. Let $\ord$ (resp. $\ordst$) be the standard (strict) order on $\SL$ and let $\join$ and $\meet$ denote the max and min of a set of integers. 
Variables $\sla,\sla',\tau_1,\ldots$ will be used to denote levels.
Variables $\slain$ and $\slaout$ will be used to denote the innermost and outermost levels.
The innermost (resp. outermost) level is the level of the innermost (resp. outermost) while loop under consideration. Outside of loops, $\slain$ and $\slaout$ are set to $\levela$.
In essence, the outermost level $\slaout$ will be used to restrict declassification, whereas the innermost level $\slain$ will restrict the type of the admissible operators in the corresponding statement.

An operator $\op$ will be associated to \emph{functional levels} of the shape $\sla_1 \to \ldots \to \sla_{ar(\op)+1}$. They can be viewed as a tuples in $\SL^{ar(\op)+1}$.

The type system uses two kinds of typing environments: 
\begin{itemize}
\item a \emph{variable typing environment} $\typenv$ is a finite maps in $\Var \to \SL$; 
\item an \emph{operator typing environment} $\typop$ maps each operator $\op \in \Operator$ to a finite map $\Delta(\op) \in \SL^2 \to\mathcal{P}(\SL^{ar(\op)+1})$.  
\end{itemize}
Intuitively, given an operator $\op$ applied in a context of  innermost level $\slain\in \SL$ and an outermost level $\slaout \in \SL$, $\typop(\op)(\slain,\slaout)$ is the set of functional levels of the shape $\sla_1 \to \ldots \to \sla_{ar(\op)+1}$ that are admissible in the caller context. See Example~\ref{ex:bubble} for an illustrating use of typing environments.
Given a typing environment $\typenv$ (resp. $\typop$), $\dom(\typenv)$ (resp. $\dom(\typop)$) denotes the domain of $\typenv$ (resp. $\typop$).

\subsubsection*{Typing Judgments and Typing Rules.} {Expression/Statement typing judgments} are of the shape $\pbl b : \sla$, with $b \in {\tt Expr} \cup {\tt Stmt}$.
The meaning is that under the \emph{innermost level} $\slain$ and the \emph{outermost level} $\slaout$, the level of $b$ is $\sla$.

The typing rules are detailed in Figure~\ref{fig:t1ts}.
Rule (OP) restricts the functional levels of the operator depending on the current innermost and outermost levels.
Rule (DCL) gives the possibility of declassifying the value of an expression of level $\sla_1$ to any level less or equal to $\slaout$.
As we have seen in the semantics, the value $\e_1$ is not declassified as is, only its length (in unary) is transmitted and the $\slaout$ compatibility is ensured by bounding this value by the second argument $\e_2$.
Those constraints will ensure that the set of possible values for variables of level $\slaout$ will not be exponential in their size.
In effect, it prevents the first dangerous code sample of Listing~\ref{ex:exp1} in the introduction.
The other rules enforce the noninterference discipline, preventing data to flow from a lower level to a strictly higher level inside loops.
Notice that it is possible to subtype statements, that is consider them to be of higher level.
On the other hand, inside a loop (i.e.,  when  $\slaout \neq \levela$), it is possible to put values of higher level into variables of lower level in Rule (ASG). This is the reason why we do not explicitly add subtyping for expressions. Notice that there are no constraints on flows outside loops (when $\slaout= \levela$) as there are no complexity concerns.
Rule (WI) sets the innermost and outermost levels $\slain$ and $\slaout$ of an outermost while loop, rule (WH) propagates the outermost level $\slaout$ and sets the new innermost level $\slain$ for nested loops.
Rule (BRK) requires  a break statement to be controlled by an expression of level at least $\slain$ to enforce the noninterference policy.

\subsubsection*{Safe Programs.}
To ensure soundness, we put restrictions on admissible operator types, depending on their computational power.

\begin{definition}\label{def:sote}
An operator typing environment $\typop$ is \emph{safe} if for each $\op \in \dom(\typop)$, $\op$ is either neutral, positive, or polynomial, $\sem{\op}$ is a polynomial time computable function, and for each $(\sla_{in},\sla_{out}) \in \SL^2$, and each $\sla_1 \to \ldots \to \sla_{ar(\op)+1} \in \typop(\op)(\sla_{in},\sla_{out})$, it holds that:
\begin{enumerate}
\item if $\op$ is a neutral operator then
$\sla_{ar(\op)+1} \ord \meet_{i=1}^{ar(\op)} \sla_i$; 
\item  if $\op$ is a positive operator then 
$\sla_{ar(\op)+1} \ord \meet_{i=1}^{ar(\op)} \sla_i$  and either $\sla_{ar(\op)+1} \ordst \slain$ or $\sla_{ar(\op)+1} = \levela$;
\item  if $\op$ is a polynomial operator then $\slaout=\levela$.
\label{sotethird}
\end{enumerate}
\end{definition}
The intuition is as follows. An operator typing environment is safe if the flow cannot go in the wrong direction for neutral and positive operators ($\sla_{ar(\op)+1} \ord \meet_{i=1}^{ar(\op)} \sla_i$); for positive operators, the computed output cannot be reused in the calling loop ($\sla_{ar(\op)+1} \ordst \slain$ or $\sla_{ar(\op)+1} = \levela$); for polynomial operators, the outermost level is enforced to be $\levela$. Hence, it is not called in a loop.

\begin{figure*}[t]
\hrulefill

\[\scalebox{1}{\begin{prooftree}
\hypo{\typenv(\mathtt{len1})=\levelb}
\hypo{\typenv(\mathtt{len})=\levela}
\infer1[(VAR)]{\typenv, \typop \vdash_{\levela}^{\levela}\mathtt{len}:\levela}
\infer2[(ASG)]{\typenv, \typop \vdash_{\levela}^{\levela}\mathtt{len1}\asg\mathtt{len}:\levelb}
\hypo{\levelb\to\levelb \in \typop(>\levela)(\levelb, \levelb)}
\hypo{\typenv(\mathtt{len1})=\levelb}
\infer1[(VAR)]{\typenv, \typop\vdash_{\levelb}^{\levelb}\mathtt{len1}:\levelb}
\infer2[(OP)]{\typenv, \typop \vdash_{\levelb}^{\levelb}\mathtt{len1}>0:\levelb}
\hypo{\romanoframedeux{$\rho$}}
\infer2[(WI)]{\typenv, \typop\vdash_{\levela}^{\levela}\whar{\mathtt{len1}>0}{\cmd 1}:\levelb}
\infer2[(SEQ)]{\typenv, \typop \vdash_{\levela}^{\levela}\mathtt{len1}\asg\mathtt{len}\sep\whar{\mathtt{len1}>0}{\cmd 1}: \levelb}
\end{prooftree}}\]
\\
\[\rho \triangleq \romanoframe{\scalebox{0.87}{\begin{prooftree}
\hypo{\vdots}
\hypo{\typenv(\mathtt{len2})=\levelb}
\hypo{\typenv(\mathtt{len})=\levela}
\infer1[(VAR)]{\typenv, \typop\vdash_{\levelb}^{\levelb}\mathtt{len}:\levela}
\hypo{\typenv(\mathtt{list})=\levelb}
\infer1[(VAR)]{\typenv, \typop\vdash_{\levelb}^{\levelb}\mathtt{list}:\levelb}
\infer2[(DCL)]{\typenv, \typop \vdash_{\levelb}^{\levelb}\declass(\mathtt{len},\mathtt{list}):\levelb}
\infer2[(ASG)]{\typenv, \typop \vdash_{\levelb}^{\levelb}\mathtt{len2}\asg\declass(\mathtt{len}, \mathtt{list}):\levelb}
\hypo{\vdots}
\hypo{\vdots}
\infer2[(WH)]{\typenv, \typop \vdash_{\levelb}^{\levelb}\whar{...}{...}:\levelb}
\infer2[(SEQ)]{\typenv, \typop \vdash_{\levelb}^{\levelb}\mathtt{len2}\asg\declass(\mathtt{len},\mathtt{list})\sep\whar{\ldots}{\ldots}:\levelb}
\infer1[(SEQ)]{\ldots}
\infer2[(SEQ)]{\typenv, \typop \vdash_{\levelb}^{\levelb}\cmd 1
:\levelb}
\end{prooftree}}}\]

\hrulefill
\caption{Typing derivation of program bubble (Listing~\ref{bubble})}\label{fig:extd}
\end{figure*}

\begin{definition}[Safety]
Given a safe operator typing environment $\typop$, a program $\prog$ is $\typop$\emph{safe} if there are a level $\sla$, a variable typing environment $\typenv$ such that  $\typenv, \typop \vdash_{\levela}^{\levela} \body(\prog): \sla$ holds. Let $\typop\saf$ be the set of $\typop$safe programs. 

A program $\prog$ is \emph{safe} if there exists a safe operator typing environment $\typop$ such that $\prog$ is $\typop$safe. 
Let $\saf$ be the set of safe programs. 
\end{definition}


\begin{example}\label{ex:bubble}
The \verb!bubble! program from Listing~\ref{bubble} can be typed with $\typenv,\typop$ such that $\typenv(\texttt{list})=\typenv(\texttt{list1})=\typenv(\texttt{len1})=\typenv(\texttt{len2})=\levelb$ and $\typenv(\texttt{list2})=\typenv(\texttt{len})=\typenv(\texttt{x})=\typenv(\texttt{y})=\typenv(\texttt{r})=\levela$, and
$$\begin{array}{rcl}
\{\levelb\to\levelb\to\levelb\} & \subset& \typop(=)(\levelb,\levelb)\cap\typop(>)(\levelb,\levelb)\cap \typop(\neq)(\levelb,\levelb),\\
\{\levelb\to\levelb\}          & \subset& \typop(\texttt{tl})(\levelb,\levelb)\cap\typop(-\levelb)(\levelb,\levelb),\\
\{\levela\to\levela\to\levela\} & \subset& \typop(+)(\levelb,\levelb),\\
\{\levela\to\levela\}          & \subset& \typop(+\levelb)(\levelb,\levelb)\cap\typop(\texttt{hd})(\levelb,\levelb)\cap\typop(\texttt{tl})(\levelb,\levelb).
\end{array}$$
\texttt{bubble} is safe:
its typing derivation tree is sampled in Figure~\ref{fig:extd}, presenting the typing of the main loop preceded by the assignment of its guard variable \verb:len1:. In Figure~\ref{fig:extd}, the statement $\cmd_1$ is the body of the loop (line 11-28 in Listing~\ref{bubble}). In Listing~\ref{bubble}, we have also provided the typing of each expression as superscript notations.
\end{example}

\subsection{Aperiodicity}
Safety is not enough on its own to ensure polytime  soundness since the declassification of one bit of information in a loop can lead to an exponential behavior as illustrated by the example of Listing~\ref{ex:exp2}.
In order to avoid this phenomenon, we complement safety with a fairly simple and natural notion of aperiodicity.

Given an expression $\e$, we define the set of \emph{undeclassified variables} of $\e$, ${U}(\e)$, by structural induction on expressions 
\begin{align*}
{U}(\x)&\triangleq \{\x\},  & {U}(\op(\bar{\e}))&\triangleq \underset{i=1}{\overset{\ell(\bar\e)}{\cup}} {U}(\e_i), \\
{U}(\declass(\e_1, \e_2))&\triangleq  {U}(\e_2). && & 
\end{align*}
Two stores $\store$ and $\store'$ are \emph{$\e$-equivalent}, noted $\store \equiv_\e \store'$, if $\forall \x \in {U}(\e)$, $\store(\x)=\store'(\x)$.

\begin{definition}[Aperiodicity]
\label{def:ap}
A program $\prog$ is \emph{aperiodic} if for each store $\store$ there are no $\pi_{(\store',\cmd')}$, $\pi_{(\store'',\cmd')}\sqsubseteq \pi_{(\store,\body(\prog))}$, such that 
\[
\cmd' = \whar{\e}{\cmd''},\quad \pi_{(\store'',\cmd')}   \sqsubset  \pi_{(\store',\cmd')},\quad \text{and}\qquad \store' \equiv_\e \store''.
\]
Let $\ap$ be the set of aperiodic programs.
\end{definition}
 In other words, for any input store of an aperiodic program, there cannot be a subtree of root $(\store',\whar{\e}{\cmd''})$ with a strict subtree of root $(\store'',\whar{\e}{\cmd''})$ such that the stores $\store'$ and $\store''$ match on undeclassified variables.
Hence, this notion of aperiodicity is very natural as it means that a given while loop is never evaluated twice under ($\e$-)equivalent stores.
Declassified variables in $\e$ are treated differently as the domain of their value is bounded by the declassification semantics. 
Hence there is no need to require aperiodicity on them for the result to hold. This can be done, but it would reduce the expressive power of the analysis.

\begin{example}
Consider the dangerous code sample from Listing~\ref{ex:exp2} in the introduction. This program is safe by taking the variable typing environment $\typenv$ defined by $\typenv(\x)\triangleq \levelb$ and $\typenv(\y)\triangleq \levela$.
However, it is not aperiodic as the undeclassified $\x$ will have value $1$ more than once in the guard of line 3.
\end{example}

\begin{example}
The program \verb!bubble! of Listing~\ref{bubble} is aperiodic.
\end{example}

\subsection{Properties of Safe and Aperiodic Programs}
\subsubsection*{Polytime Soundness and Completeness.}

A program $\prog$ of the shape $\main(\bar{\x})\{\cmd\ \ret \y\}$ computes a partial function $\sem{\prog}: \W^{\ell(\bar{\x})} \to \W$ defined as 
\[
\sem{\prog}(\bar{\w}) = \v\text{ iff }(\store[\bar{x}\leftarrow\bar{\w}], \cmd) \toe (\top, \store')\text{ and }\store'(\y)=\v.
\]
If $\sem{\prog}$ is a total function, we say that the program is \emph{terminating}. Let $\TERM$ be the set of terminating programs. Given a set of programs $S$, we define $\sem{S}$ by $\sem{S} \triangleq \{ \sem{\prog} \ | \ \prog \in S\}$.

We now show a completeness result stating that for each function $f$ in $\FP$, the class of functions computable in polynomial time by a Turing machine, there exists a safe and aperiodic program $\prog$ such that $f=\sem{\prog}$. This completeness result relies on simulating any Turing machine that halts in polynomial time with a safe and aperiodic terminating program.
\begin{restatable}[Completeness]{theorem}{thmfpcomplete}\label{thm:fpcomplete}
$\FP \subseteq \sem{\saf \cap \ap}$.
\end{restatable}

For soundness, the intuition is as follows. 
For a fixed loop of level $\sla$, the safety property implies a polynomial bound on the set of potentially computed values for variables of level greater than or equal to $\sla$.
Aperiodicity then makes it impossible to reach twice the same configuration of variables of level at least $\sla$.
Hence the number of iterations is polynomially bounded. Moreover, variables of strictly lower level can only increase polynomially by a constant (using positive operators). Repeating the reasoning for any level, the total runtime is polynomially bounded in the program input. This bound is obtained by a constant composition of polynomials.

\begin{restatable}[Soundness]{theorem}{thmpoli}\label{thm:poli}
$ \sem{\saf \cap \ap}\subseteq \FP $.
\end{restatable}
As a corollary, this implies termination. Indeed, on any input store, the space of reachable values (and hence reachable stores) corresponding to the program derivation tree is polynomially bounded.

\begin{corollary}[Termination]
$ \sem{\saf \cap \ap}\subsetneq \TERM $.
\end{corollary}

Safe and aperiodic programs contain polytime programs with loops guarded by data that may increase.
For example, the program $\tt bubble$ of Listing~\ref{ex:t1} has an inner loop controlled by $\tt len2$, which is reinitialized at the size of $\tt len$ at each iteration of the outer loop.
Such programs cannot be captured by existing criteria, e.g.,~\cite{M11}. Consequently, this result improves greatly on the expressive power of noninterference-based type systems for complexity.

\subsubsection*{Complexity of Type Inference and Aperiodicity.}
We now show that type inference can be decided in polynomial time in the \emph{size of a program} $\size{\prog}$ (i.e., number of symbols) , which implies that type inference can be efficiently implemented.
\begin{restatable}[Type inference]{theorem}{thmtiproc}\label{thm:tiproc}
Given a safe operator typing environment $\typop$, deciding whether $\prog \in \typop\saf$ can be done in time $\mathcal{O}(\size{\prog}^3)$.
\end{restatable}

Now we show that aperiodicity  is a $\Pi_1^0$-complete problem in the arithmetical hierarchy.

\begin{restatable}[Aperiodicity]{theorem}{thmap}\label{thm:ap}
Deciding whether a program  $\prog$ is aperiodic (i.e., $\prog \in \ap$) is $\Pi^0_1$-complete.
\end{restatable}

This undecidability result is not a negative result as the set of programs in the class $\FP$ is known to be $\Sigma^0_2$-complete~\cite{H79}. Hence aperiodicity is a strictly simpler problem. Moreover, the termination hypothesis in previous work (e.g., ~\cite{M11,HP18}) is $\Pi^2_0$-complete (~\cite{EGSZ11}), thus strictly harder than aperiodicity as a consequence of Post's Theorem.

\subsubsection*{Sound and Complete Criterion with Decidable Aperiodicity.}
Last but not least, it is possible to specify a decidable aperiodicity condition that preserves the completeness of Theorem~\ref{thm:fpcomplete}. 
We define a for loop statement $\ifor \x = \e \ito \d\ \{ \cmd \}$ as syntactic sugar for the statement $\x \asg \d \sep \while(\x \geq \e)\{\cmd \sep \x \asg \x -1\}$, under the proviso that $\x$ does not occur in $\cmd$.
Let $\emph{for-programs}$ be programs that only contains for loop (and no other while loop). For-programs are trivially aperiodic, by definition. Let $\forp$ be the set of for-programs.
We can show that safe for-programs are sound and complete for $\FP$.

\begin{theorem}[Decidable criterion]\label{thm:for}
$ \sem{\saf \cap \forp } =\FP$ 
\end{theorem}

Soundness and tractability of type inference are obtained as a direct corollary of Theorems~\ref{thm:poli} and~\ref{thm:tiproc} as for-programs are aperiodic programs. Completeness is also obtained easily as the programs used in the proof of Theorem~\ref{thm:fpcomplete} are all for-programs.

\section{Application Example: Characterizing the Class BFF}
We now enrich the language introduced in Section~\ref{sec:t1} with \emph{procedures},  \emph{oracle calls}, and \emph{closures} with the aim of characterizing the second-order complexity class of Basic Feasible Functionals, $\BFF$.
The language additions are minimal and aim at extending the language to second-order -- a prerequisite to characterize $\BFF$ -- and are in no way linked to noninterference or declassification issues. 

\subsection{Type-2 Programming Language}
\label{s:prog}

\begin{figure}[t]
\hrulefill
$$
\begin{array}{llll}
\texttt{Expr}          \qquad & \e
  & \rgl &\x
   \ |\ \op(\bar{\e})  \ | \ 
   \declass(\e,\e)\ | \   \X (\bar{\e}) \\
\texttt{Stmt}                  &  \cmd 
    & \rgl &\skp   \ | \  \x  \asg \e   \ | \  \cmd \sep \cmd \ |\  \ifa (\e) \{\cmd\} \elsea \{\cmd\} \ |\ \\
   & &  &\while(\e)\{ \cmd \}  \ | \ \brk(\e)\\
  \texttt{Proc}                & \procn    & \rgl & \proc(\bar{\X},\bar{\x})\{[\var \bar{\y}\sep\!\!]\ \cmd  \ \ret \x\} \\
      \texttt{Terms}                 & \term   & \rgl & \x  \ | \   
      \callcc\proc(\bar{\clos},\bar{\term}) \\
      \texttt{Closures}              & \clos   & \rgl & \X \ | \ \abs{\bar{\x}}.\term \\
       \texttt{Prg}             & \prog 
  & \rgl &  \term \ | \ \letin{{\procn}} \prog \ | \ \boite{[\va]}{\prog} 
 \end{array}
 $$
 \hrulefill
\caption{Syntax of second-order programs}
\label{fig:synt}
\end{figure}

\subsubsection*{Syntax.}

\begin{figure*}[t]
\begin{minipage}[t]{0.45\textwidth}
\begin{lstlisting}[basicstyle=\footnotesize\ttfamily,firstnumber=2,caption={Procedure \texttt{J}}, captionpos=b,  label={ex:J}]
$\niceinstr{declare} \verb!J!$($\Y$,l,m,n){
  $\var$b$\sep \var$t$\sep \var$z$\sep \var$l0$\sep\var$n0$\sep$
  l0$\asg$l$\sep$
  n0$\asg$n$\sep$
  z$\asg$$\varepsilon$$\sep$
  b$\asg$n+m+1$\sep$
  $\while$(n>0){
    $\brk$(|$\Y$(l,n)|>|$\Y$(l0,n0)|)$\sep$
    t$\asg$$\truncate$($\Y$(l,n),b)$\sep$
    l$\asg$$\gauche$($\droite$(t))$\sep$
    n$\asg$$\declass(\droite$($\droite$(t)),b)
  }$\sep$
  $\ifa$(n=0){
    z$\asg$l
  }
  $\ret$z
}
\end{lstlisting}
\end{minipage}
\begin{minipage}{0.1\textwidth}
	\phantom{m}
\end{minipage}
\begin{minipage}[t]{0.4\textwidth}
\begin{lstlisting}[basicstyle=\footnotesize\ttfamily,firstnumber=19, caption={Procedure \texttt{K}}, captionpos=b, label={ex:K}]
$\niceinstr{declare} \verb!K!$($\X$,p,q,r,s){
  $\var$i$\sep \var$j$\sep$
  i$\asg$r$\sep$
  $\while$(s>0){
    $\brk$(|$\X$(i)|>|$\X$(r)|)$\sep$
    i$\asg$$\truncate$($\X$(i),p)$\sep$
    s$\asg$s-1
  }$\sep$
  j$\asg$pad(p+q+1,$\pair$(i,s))
  $\ret$j
}
\end{lstlisting}
\def\createlinenumber#1#2{
    \edef\thelstnumber{%
        \unexpanded{%
            \ifnum#1=\value{lstnumber}\relax
              #2%
            \fi}%
        \ifx\thelstnumber\relax\else
        \expandafter\unexpanded\expandafter{\thelstnumber}%
        \fi
    }
}
\bgroup
\let\thelstnumber\relax
\createlinenumber{1}{1}
\createlinenumber{2}{2-18}
\createlinenumber{3}{19-29}
\createlinenumber{4}{30}

\begin{lstlisting}[basicstyle=\footnotesize\ttfamily,caption={Program \texttt{I}},label={fig:bruce},captionpos=b]
$\niceinstr{box}[\F,\text{u},\text{v},\text{w}]\ \niceinstr{in}$
$\niceinstr{declare} \verb!J!$($\Y$,l,m,n){...} $\niceinstr{in}$
$\niceinstr{declare} \verb!K!$($\X$,p,q,r,s){...} $\niceinstr{in}$
$\callcc \verb!J!$($\abs{\!\!\text{x},\!\text{y}}.\callcc \verb!K!$($\F$,v,w,x,y),u,v,w)
\end{lstlisting}
\egroup
\end{minipage}
\end{figure*}

\emph{Programs} are defined by the grammar of Figure~\ref{fig:synt} and can be either a term, a \emph{procedure declaration} within a program, or the declaration of a \emph{boxed variable}, called \emph{box}, followed by a program. Boxed variables represent program inputs and can be either order-$1$ upper-case variables $\X \in \Var_1$ or order-$0$ lower-case variables $\x \in \Var_0$. In a box, the variable $\va \in \Var \triangleq \Var_0 \uplus \Var_1$ can be of arbitrary order.

\emph{Terms} $\term$ are order-$0$ constructs and can be either a variable $\x$ 
or a \emph{procedure call} $\callcc\proc(\bar{\clos},\bar{\term})$ corresponding to the application of procedure of name $\proc$ to order-$1$ inputs $\bar{\clos}$ and order-$0$ inputs $\bar{\term}$.

A \emph{procedure declaration} $\proc(\bar{\X},\bar{\x})\{[\var \bar{\y}\sep\!]\ \cmd  \ \ret \x\}$ is the corresponding order-2 abstraction. It maps order-$1$ and order-$0$ \emph{parameters} $\param(\proc) \triangleq \{\bar{\X},\bar{\x}\}$ to a order-$0$ output $\x$ and consists of a body statement $\body(\proc)\triangleq \cmd $ with optional declaration of \emph{local variables} $\local(\proc)\triangleq\{\bar{\y}\}$. For convenience, we identify a procedure declaration with its procedure name. Moreover, we assume that each order-$1$ procedure parameter comes with a fixed arity $ar(\X)$.

\emph{Statements} are the same as in Figure~\ref{fig:t1syn} and thus can still only assign to order-$0$ variables.
Similarly, \emph{expressions} include those of Figure~\ref{fig:t1syn} plus a new construct $\X(\bar{\e})$, named \emph{oracle call}, consisting in the application of a order-$1$ variable $\X$ to some sequence of expressions $\bar{\e}$, called the \emph{input data}. We assume that they are always fully applied, i.e., that $\ell(\bar{\e})=ar(\X)$.

 \emph{Closures} are order-$1$ abstractions corresponding to oracle calls and can be written 
 either as an order-$1$ variable $\X$ or 
 as $\abs{\bar{\x}}.\term$, where the order-$0$ term $\term$ may contain free variables. Notice that procedure calls can only take closures as order-$1$ parameters and that procedure declarations with no oracle call are exactly the first-order programs of Section~\ref{sec:t1}. 
 

A variable is free if it is neither boxed nor bound by one of the two possible abstractions.  A program is \emph{closed} if it has no free variable.
We define the following syntactic sugar:
\begin{align*}
\letin{\bar{\procn}}\prog&\triangleq \letin{\procn_1}\ldots\letin{\procn_n}\prog\\
\boite{[\bar \va]}{\prog}&\triangleq \boite{[\bar \va_1]}{\ldots \boite{[\bar \va_m]}{\prog}}
\end{align*}
with $\bar{\procn} \triangleq \procn_1,\ldots,\procn_n$ and $\bar{\va}\triangleq \bar \va_1,\ldots,\bar\va_m$.

Throughout the rest of the paper, we will restrict attention to closed  programs in \emph{normal form}. These consist of programs that can be written as follows $\boite{[\bar{\X},\bar{\x}]}{\letin{\bar{\procn}} \term},$
 for some term $\term$ such that the following \emph{well-formedness conditions} hold:
\begin{inparaenum}[i)]
\item \emph{no name clash}: the set of the term free variables, each set of procedure parameters, and each set of procedure local variables are all pairwise distinct,
\item  \emph{closed procedures}: each variable of a procedure body is either a procedure parameter or a local variable,
\item \emph{determinism}: each procedure name called in $\term$  is declared exactly once in $\procn$.
\end{inparaenum}
A program in normal form computes a second-order functional.

\begin{example}\label{ex:bruce}
Program \verb!I! in Listing~\ref{fig:bruce} is closed and in normal form and computes  a variant of Cook-Urquhart's second-order bounded iterator~\cite{CU93}
\begin{align*}
\mathcal{I}& :  (\W \to \W) \times \W^3 \to \W \\
\mathcal{I}(f,\u,\v,\w) &\triangleq (\lambda x.\!\! \upharpoonright\!\!(f(x),\v))^{\size{\w}}(\u),
\end{align*}
where $f^n$ is the $n$-iterate of $f$ and $\upharpoonright \ : \W \times \W \to \W$, defined by
\[
\upharpoonright\!\!(\v ,\w)\triangleq \begin{cases} \v &\text{if }\size{\v} \leq \size{\w} \\ \v' \text{ s.t. } \size{\v'} =\size{\w} \wedge \v =\v'\cdot\v'' & \text{otherwise,}\end{cases}
\]
is a function truncating its first operand with respect to the size of its second operand.

In this example, words can encode lists to represent pairs and triples. To account for this, the alphabet $\Sigma$ on which words are defined includes a constructor symbol $\#$ in addition to $1$ and $0$.
On boolean and arithmetic operators, the function $\sem{\op}$ behaves as expected on suitable data and returns the empty word $\epsilon$ in case of failure, e.g., $\sem{\sz}(100\#11)=110$, $\sem{\leq}(110,111)=1$, and $\sem{+1}(1\#0)=\epsilon$. 
Constructor and destructor semantics function can be defined similarly: $\sem{\pair}(\u,\v)\triangleq  \u\cdot\#\cdot\v$, if $ \# \not\in \u$, $\sem{\pair}(\u,\v)\triangleq \epsilon$, otherwise, 
$\sem{\gauche}(\sem{\pair(\u,\v)})\triangleq \u$ and $\sem{\droite}(\sem{\pair(\u,\v) }) \triangleq \v$.
The operator $\pad$ is a padding operator defined by: $\sem{\pad}(\u,\v)\triangleq \sem{\pair(\v,0^n)}$, if $ \exists n \geq 0,\ \size{\v} + n+1 = \size{\u}$, and $\sem{\pad}(\u,\v)\triangleq \epsilon$, otherwise. We use this operator in procedure \verb!K! to produce outputs of non-decreasing size (line 27).
Finally, the operator $\truncate$ is the syntactic counterpart of the function $\upharpoonright$ defined by $\sem{\truncate}(\v,\w) \triangleq \upharpoonright\!\!(\v ,\w)$ and is used in the two procedures to bound the oracle outputs. 

Program \verb!I!  makes use of two intermediate procedures \verb!J!  and  \verb!K! defined in Listings~\ref{ex:J} and~\ref{ex:K}.  \verb!K! iterates at most $s$ times an oracle call $\X(\text{i})$ (hence computes $(\lambda x.\!\!\upharpoonright\!\!(f(x),\v))^{\size{s}}(\u)$), for some $s$ up to which $f$ outputs does not exceed the size of the first call to $f$. This check is performed using a break statement. Hence \verb!K!  computes a functional in $\flr$. \verb!J! iterates $\size{\w}$ times \verb!K!, using a closure. 

Procedure \verb!J! uses the $\declass$ construct. This will be required for the program to type as, in the loop, the output of an oracle call is assigned to the variable \verb!n! that guards the loop. In general, such behaviors will be prohibited as there may be exponentially many oracle calls  (even for a fixed size) unless an explicit declassification is performed. Here the declassification has the effect of restricting the output of the oracle call into a polynomially bounded domain.
\end{example}

\begin{figure*}[t]
\begin{minipage}{\textwidth}
\hrulefill
\[\begin{prooftree}
\hypo{\phantom{ \Imp \x  \asg \e\to } }
\infer1[]{(\sigma,\store,\phi \Imp \x) \toexp \store(\x)}
\end{prooftree}
\qquad
\begin{prooftree}
\hypo{(\sigma,\store,\phi,\bar{\e}) \toexp \bar{\w}}
\infer1[]{(\sigma,\store,\phi \Imp \op(\bar{\e})) \toexp \sem{\op}(\bar{\w})}
\end{prooftree}
\qquad
\begin{prooftree}
\hypo{\forall i \leq 2, \ (\sigma,\store,\phi,\e_i) \toexp \w_i}
\infer1[]{(\sigma,\store,\phi \Imp \declass(\e_1,\e_2)) \toexp 1^{\min(|w_1|, |w_2|)}}
\end{prooftree}
\]

\[
\romanobox{
\begin{prooftree}
\hypo{(\sigma,\store,\phi \Imp \bar \e) \toexp {\bar \v}}
\hypo{\phi(\X)= \abs{\bar{\x}}.\term}
\hypo{(\sigma,\store[\bar{\x} \leftarrow  \bar{\v}],\term ) \toenv \w }
\infer3[(Orc)]{(\sigma,\store,\phi \Imp \X(\bar{\e})) \toexp \w}
\end{prooftree} 
}    
\]

\[
\begin{prooftree}
\hypo{\phantom{ \Imp \x  \asg \e^O \to }  }
\infer1[]{(\sigma,\store,\phi \Imp \skp) \tost (\top,\store)}
\end{prooftree}
\qquad
\begin{prooftree}
\hypo{(\sigma,\store,\phi \Imp \e) \toexp \w }
\infer1[]{(\sigma,\store,\phi \Imp \x  \asg \e) \tost (\top, \store[\x  \leftarrow \w])}
\end{prooftree}
\qquad
\begin{prooftree}
\hypo{(\sigma,\store,\phi \Imp \cmd_1) \tost (\bot,\store') }
\infer1[]{(\sigma,\store,\phi \Imp \cmd_1 \sep \cmd_2) \tost (\bot,\store')}
\end{prooftree}
\]

\[
\scalebox{0.9}{
\begin{prooftree}
\hypo{(\sigma,\store,\phi \Imp \cmd_1) \tost (\top,\store') }
\hypo{(\sigma,\store',\phi \Imp \cmd_2) \tost (\square,\store'')}
\infer2[($\square \in \{\top,\bot\}$)]{(\sigma,\store,\phi \Imp \cmd_1 \sep \cmd_2) \tost (\square,\store'')}
\end{prooftree}
}
\]

\[
\begin{prooftree}
\hypo{(\sigma,\store,\phi \Imp \e) \toexp \w}
\hypo{(\sigma,\store,\phi \Imp \cmd_\w) \tost (\square,\store')}
\infer2[($\square \in \{\top,\bot\}, \w \in \{\underline{\false},\true\}$)]{(\sigma,\store,\phi \Imp \ifa (\e) \{\cmd_{\true}\} \elsea \{\cmd_{\underline{\false}}\} ) \tost (\square,\store')}
\end{prooftree}
\]

\[
\begin{prooftree}
\hypo{(\sigma,\store,\phi \Imp \e) \toexp \underline{\false}} 
\infer1[]{(\sigma,\store,\phi \Imp \while (\e) \{\cmd\}) \tost (\top, \store)}
\end{prooftree}
\qquad 
\begin{prooftree}
\hypo{(\sigma,\store,\phi \Imp \e) \toexp \true}
\hypo{(\sigma,\store,\phi \Imp \cmd \sep  \while (\e) \{\cmd\}) \tost (\square,\store')}
\infer2[($\square \in \{\top,\bot\}$)]{(\sigma,\store,\phi \Imp  \while (\e) \{\cmd\}) \tost (\top,\store')}
\end{prooftree}
\]

\[
\begin{prooftree}
\hypo{(\sigma,\store,\phi \Imp \e) \toexp\underline{\false}} 
\infer1[]{(\sigma,\store,\phi \Imp \brk(\e)) \tost (\top,\store)}
\end{prooftree}
\qquad
\begin{prooftree}
\hypo{(\sigma,\store,\phi \Imp \e) \toexp\true} 
\infer1[]{(\sigma,\store,\phi \Imp \brk(\e)) \tost (\bot,\store)}
\end{prooftree}
\qquad
\romanobox{
\begin{prooftree}
\hypo{\phantom{(\sigma,\store)}}
\infer1[(Var)]{(\sigma,\store,\x) \toenv \store(\x)}
\end{prooftree}
}
\]

\[
\romanobox{
\scalebox{1}{\begin{prooftree}
\hypo{\proc(\bar{\X},\bar{\x})\{ \cmd  \ \ret \z\}  \in \sigma}
\hypo{(\sigma,\store,\bar{\term}) \toenv \bar{\w}}
\hypo{(\sigma,\store[\bar{\x} \leftarrow \bar{\w},\bar{\y} \leftarrow \bar{\epsilon}],\bar{\X} \mapsto \bar{\clos},\cmd ) \tost (\square,\store') }
\infer3[($\square \in \{\top,\bot\}$)\ (Call)]{(\sigma,\store  \Imp \callcc \proc(\bar{\clos},\bar{\term})) \toenv \store'(\z)}
\end{prooftree}}
}
\]

\[
\romanobox{
\begin{prooftree}
\hypo{(\sigma\cup\{\procn\},\store,\prog) \toenv \w}
\infer1[(Dec)]{(\sigma,\store,\letin{\procn} \prog) \toenv \w}
\end{prooftree} 
} 
\qquad
\romanobox{
\begin{prooftree}
\hypo{(\sigma,\store,\prog) \toenv \w}
\infer1[(Box)]{(\sigma,\store,\boite{[\va]}{\prog}) \toenv \w}
\end{prooftree} 
}   
\]
\hrulefill
\end{minipage}
\caption{Semantics of second-order programs}
\label{fig:prog2}
\end{figure*}

\subsubsection*{Operational Semantics.}
In what follows, let $f,g,\ldots$ denote total functions in $\W \to \W$. \emph{Stores} $\store$ and \emph{environments} $\phi$ are defined by
\begin{align*}
&({\tt Stores})   &&\store \in (\Var_0 \to \W) \uplus   \Var_1 \to (\W \to \W)  \\
&({\tt Env})     &&\phi \in \Var_1 \to {\tt Closures} 
\end{align*}

Stores are total functions assigning words or functions on words to a variable, depending on its order. Environments assign a closure to an order-1 variable $\X$.  Again, $\store_\emptyset$ will denote the empty store, mapping each order-$0$ variable $\x$ to $\varepsilon$ and each order-$1$ variable $\X$ to the constant function $\lambda \w.\varepsilon$. In the same vein, $\store[\bar{\x} \leftarrow \bar{\w},\bar{\X} \leftarrow \bar{f}]$ 
denotes a store update. We now introduce the following judgments  
\begin{align*}
\toexp &\in (\mathcal{P}({\tt Proc}) \times {\tt Stores} \times {\tt Env} \times {\tt Expr}) \to \W \\
\tost &\in (\mathcal{P}({\tt Proc}) \times {\tt Stores} \times {\tt Env} \times {\tt Stmt}) \to (\{\top,\bot\} \times {\tt Stores}) \\
\toenv &\in (\mathcal{P}({\tt Proc}) \times {\tt Stores} \times {\tt Prg} ) \to \W
\end{align*}

The judgment $(\sigma,\store,\phi,\e) \toexp  \w$ means that the expression $\e$ evaluates to the word $\w \in \W$ wrt. the set of procedure declarations $\sigma$, the store $\store$, and the environment $\phi$.
The judgment $(\sigma,\store,\phi,\cmd) \tost (\square,\store')$, with $\square \in \{\top,\bot\}$, expresses that, under the set of procedure declarations $\sigma$, the store $\store$, and the environment $\phi$, the statement $\cmd$ terminates and outputs the store $\store'$. As in Section~\ref{s:declass}, the symbol $\bot$ indicates that a break instruction has been executed.
The judgment $(\sigma,\store,\prog)\toenv \w$ deterministically maps a set $\sigma$ of procedure declarations, a store $\store$, and a program $\prog$ in normal form to a word $\w \in \W$.

The operational semantics is defined in Figure~\ref{fig:prog2}. Most rules are very similar to that of Figure~\ref{fig:t1sem}. The new rules, corresponding to the new syntactic constructs, are highlighted with \romanotextbox{boxes}. We provide an intuition for each of them. Rule (Orc) deals with an oracle calls. The rules evaluates the order-$0$ term $\term$ of the closure $\abs{\bar \x}.\term$ corresponding to the order-$1$ variable $\X$ in the environment $\phi$ wrt. the values $\bar\v$ of the operands $\bar \e$.  
Rule (Var) outputs the word $\store(\x)$
. 
In Rule (Call),  $\bar{\X} \mapsto \bar{\clos}$, with $\ell(\bar{\X})=\ell(\bar{\clos})$, is the environment mapping each $\X_i \in \Var_1$ to the closure $\clos_i$. It is important to stress that Rule (Call) is the only rule that updates the environment to the closures passed as arguments: hence, oracle calls are fixed for each procedure call. 
Rule (Dec) just adds the procedure declaration to the set of procedures in the judgments and Rule (Box) is just a no-op rule.

The notation $\pi_c$ is overloaded to denote the evaluation tree of root $c$ using the rules of Figure~\ref{fig:prog2}, with $c$ being an element in the definition domain of $\toexp$, $\tost$, or $\toenv$ and the strict subtree relation $\sqsubset$ is updated accordingly.

As we will see shortly, the type discipline will ensure that a  program $\prog= \boite{[\bar{\X} ,\bar{\x}]}{\letin{\bar{\procn}}\term}$ computes the second-order partial functional $\sem{\prog}$ in $(\W \to \W)^{\ell(\bar{\X})} \to \W^{\ell(\bar{\x})} \to \W,$
defined by:
\[
\sem{\prog}(\bar{f},\bar{\w})=w \text{ iff } (\emptyset, \store_\emptyset[\bar{\x} \leftarrow \bar{\w},\bar{\X} \leftarrow \bar{f}],\prog) \toenv  \w.
\]

The store $\store_\emptyset[\bar{\x} \leftarrow \bar{\w},\bar{\X} \leftarrow \bar{f}]$ is called an \emph{input store}.
If $\sem{\prog}$ is a total function, the program $\prog$ is said to be \emph{terminating}. Let $\SN$ be the set of terminating programs.

\subsection{FLR Restrictions and Aperiodicity}
\subsubsection*{Syntactic restrictions.}
For showing soundness and completeness wrt $\BFF$, we need to put some restrictions and hypothesis on the programming language. First, we assume that the set of operators $\Operator$ includes all the operators $\{=, <, \leq, 0, +1, -1, \non, \et, \ou\}$ used in Section~\ref{s:declass} and all the operators defined in Example~\ref{ex:bruce}. Indeed, this example is used in the proof of completeness and, hence, it is important to include all its operators for completeness to hold. Now we syntactically restrict the study to \emph{guarded programs}. 
\begin{definition}[Guarded program]
A program is \emph{guarded} if  
\begin{enumerate}
\item oracle calls cannot be nested and appear either in assignments or in break statements $\brk(\size{\X(\bar{\e})}>\size{\X(\bar{\x})})$,\label{c1}
\item in a loop, each assignment containing an oracle call $\X(\bar{\e})$ is preceded by a break statement $\brk(\size{\X(\bar{\e})}>\size{\X(\bar{\x})})$. \label{c2}
\end{enumerate}
\end{definition}

Guardedness just put simple syntactic restrictions on how oracle calls can be used in a program.
The intuition behind is to ensure the $\flr$ property, i.e., that size of oracle calls cannot increase more than a constant number of times during the program execution. The aim of item~(\ref{c2}) is to ensure $\flr$ by the increase of oracle outputs with a break statement before assigning them. The restriction of oracle calls to break statements and assignments~(\ref{c1}) avoids uncontrolled increases in other expressions of the program (guard of while, of conditionals, ...). However, this syntactic restriction is not sufficient on its own to ensure that programs compute a functional in $\flr$ (a fortiori in $\spt$). This property will be achieved by complementing the restriction by the typing discipline. In what follows, we will restrict our study to guarded programs.

\begin{example}
Program \verb!I! of Listing~\ref{fig:bruce} is guarded.
Indeed, the oracle calls line 10 in procedure \verb!J! and at line 24 in procedure \verb!K! are immediately preceded by a $\brk$ statement of the good shape (\ref{c2}). Moreover, there are no other oracle calls in the program (\ref{c1}).
\end{example}

\subsubsection*{Aperiodicity Revisited.}

We lift the notion of aperiodicity introduced in Definition~\ref{def:ap} to this second-order programming language. The set of \emph{undeclassified variables} of $\e$, ${U}(\e)$, is extended by
\[
{U}(\X(\bar{\e}))\triangleq \underset{i=1}{\overset{\ell(\bar\e)}{\cup}}{U}(\e_i),
\]
and the definition of $\equiv_\e$ is updated accordingly.

\begin{definition}[II-Aperiodicity]\label{coro:ap}
A program $\prog$ is \emph{II-aperiodic} if, for each store $\store$, there are no $\pi_{(\sigma',\store',\phi',\cmd)}$ $\pi_{(\sigma'',\store'',\phi'',\cmd)}\sqsubseteq \pi_{(\emptyset,\store,\prog)}$ s.t. 
\[
\cmd = \whar{\e}{\cmd'},\quad \pi_{(\sigma',\store',\phi',\cmd)} \sqsubset \pi_{(\sigma'',\store'',\phi'',\cmd)},\ \text{and}\quad  \store' \equiv_\e \store''.
\]
Let $\dap$ be the set of II-aperiodic programs. 
\end{definition}
II-aperiodicity is very similar to the notion of aperiodicity for first-order programs in Definition~\ref{def:ap}:
it still enforces that there cannot be two consecutive calls to a while loop of a given procedure under equivalent stores.

For a given set of programs $S$, let $\sem{S}\triangleq \{ \sem{\prog} \ | \ \prog \in S\}$ denote the set of functions computed by programs in $S$.  E.g., $\sem{\ASN}$ is the set of functions computed by aperiodic and terminating programs. In aperiodic and terminating programs, termination does not depend on the oracle output.

\begin{example}\label{ex:sns}
The program \verb!I! of Listing~\ref{fig:bruce} is in $\SN$.
Indeed, procedure \verb!K! trivially terminates on any input.
Moreover, when called on the appropriate closure (line 30 of Listing~\ref{fig:bruce}), procedure \verb!J! assigns to \verb!n! a value that has decreased by at least $1$ or that is equal to $0$ (line 8 of Listing~\ref{ex:J}).
Notice however that it does not entail that the procedures are terminating.
Indeed, procedure \verb!J! only terminates on specific closures (including $\abs{\x,\y}.\callcc \verb!K!(\F,\variable{v},\variable{w},\x,\y)$).
Program \verb!I! is also aperiodic: on any input store, in procedure \verb!K!, the value of variable \verb!s! used in the guard of the loop  strictly decreases and, in procedure \verb!J!, the value of variable \verb!n! used in the guard of the loop strictly decreases, as this procedure is called on the closure $\abs{\x,\y}.\callcc \verb!K!(\F,\variable{v},\variable{w},\x,\y)$.
We conclude that \verb!I! $\in \ASN$.
\end{example}

As the proof of Theorem~\ref{thm:ap} is only based on procedure bodies,  II-aperiodicity is also $\Pi_0^1$-hard and in $\Pi^1_1$.

\begin{corollary}
Deciding whether a program $\prog$ is II-aperiodic (i.e., $\prog \in \dap$) is $\Pi^0_1$-hard.
\end{corollary}

\begin{figure*}[p]
\begin{minipage}{\textwidth}
\hrulefill
\[
\scalebox{1}{
\begin{prooftree}
\hypo{\typenv(\x)=\sla}
\infer1[
]{\pbl \x: \sla}
\end{prooftree}
\qquad
\begin{prooftree}
\hypo{\sla_1 \to \cdots \to \sla_{ar(\op)+1} \in \Delta(\op)(\sla_{in},\sla_{out})}
\hypo{\forall i \leq ar(\op),\ \pbl \e_i: \sla_i}
\infer2[
]{\pbl  \op(\bar{\e}): \sla_{ar(\op)+1+1} }
\end{prooftree}
}
\]

\[
\scalebox{1}{
\begin{prooftree}
\hypo{\pbl \e_1: \sla_1}
\hypo{\pbl \e_2: \sla_{out}}
\hypo{\sla_1 \ord \sla\ord \sla_{out}}
\infer3[
]{\pbl  \declass(\e_1,\e_2): \sla}
\end{prooftree}
}
\qquad
\scalebox{1}{\romanobox{\begin{prooftree}
\hypo{\forall i \leq \ell(\bar{\e}),\ \pbl \e_i : \sla_i}
\infer1[(ORC)]{\pbl \X(\bar{\e}) : \infty}
\end{prooftree}}}
\]

\[
\scalebox{1}{
\begin{prooftree}
\hypo{\pbl  \cmd\ : \sla_1}
\hypo{\sla_1 \leq \sla_2}
\infer2[
]{\pbl  \cmd\ : \sla_2}
\end{prooftree}
}\]

\[
\begin{prooftree}
\hypo{\phantom{\romanobox{$\sla_2 \neq \infty$}}}
\infer1[
]{\pbl \skp\ : \levela}
\end{prooftree}
\qquad
\scalebox{1}{
\begin{prooftree}
\hypo{\typenv(\x) = \sla_1}
\hypo{\pbl \e: \sla_2}
\hypo{(\sla_{out} = \levela) \vee (\sla_1  \ord \sla_2)}
\hypo{$\romanobox{$\sla_2 \neq \infty$}$}
\infer4[(ASG)
]{\pbl \x \asg \e \ : \sla_1}
\end{prooftree}}
\]

\[
\scalebox{1}{
\begin{prooftree}
\hypo{\pbl  \cmd_1 :\sla}
\hypo{\pbl \cmd_2:\sla}
\infer2[
]{\pbl \cmd_1 \sep \cmd_2 \ :\sla}
\end{prooftree}}
\qquad
\scalebox{1}{
\begin{prooftree}
\hypo{\pbl \e: \sla}
\hypo{\pbl \cmd_{\true}: \sla}
\hypo{\pbl \cmd_{\false}:\sla}
\infer3[
]{\pbl \ifa (\e) \{\cmd_{\true}\} \elsea \{\cmd_{\false}\}\ :\sla}
\end{prooftree}
}
\]

\[
\scalebox{1}{
\begin{prooftree}
\hypo{\typenv,\typop \vdash^{\sla}_\sla \e: \sla}
\hypo{\typenv,\typop \vdash^\sla_\sla \cmd : \sla}
\hypo{\levelb \ord \sla}
\hypo{$\romanobox{$\sla \neq \infty$}$}
\infer4[(WI)
]{\typenv,\typop \vdash^{\levela}_\levela  \while (\e) \{\cmd\}\ :\sla}
\end{prooftree}
}
\]

\[
\scalebox{1}{
\begin{prooftree}
\hypo{\typenv,\typop \vdash^\sla_{\sla_{out}} \e: \sla}
\hypo{\typenv,\typop \vdash^\sla_{\sla_{out}} \cmd : \sla}
\hypo{\levelb \ord \sla \ord \sla_{out}}
\hypo{$\romanobox{$\sla \neq \infty$}$}
\infer4[(WH)
]{\pbl  \while (\e) \{\cmd\}\ :\sla}
\end{prooftree}
}
\qquad
\scalebox{1}{
\begin{prooftree}
\hypo{\pbl \e  : \sla}
\hypo{\sla_{in} \ord \sla}
\infer2[
]{\pbl \brk(\e)  : \sla_{in}}
\end{prooftree}
}
\]

\[
\scalebox{1}{\romanobox{\begin{prooftree}
\hypo{\pbl \X(\bar{\e}) : \infty}
\hypo{\pbl \X(\bar{\x}) :\infty}
\hypo{\forall i \leq \ell(\bar{\x}),\ \pbl \x_i : \sla_i}
\hypo{\sla_{out} \ordst \meet_i \sla_i}
\infer4[(OBK)]{\pbl \brk(\size{\X(\bar{\e})} >\size{\X(\bar{\x})})  : \sla_{in}}
\end{prooftree}
}}\]
%
\dotfill
%
\[
\romanobox{
\scalebox{1}{
\begin{prooftree}
\hypo{\pbla \bar{\X} : \overline{\bar{\TW} \to \TW}}
\hypo{\pbla \bar{\x}: \bar{\TW}}
\hypo{\pbla \bar{\y} : \bar{\TW}}
\hypo{\pbla \x : {\TW}}
\infer4[(DEF)]{\pbla \proc(\bar{\X},\bar{\x})\{[\var \bar{\y}\sep\!]\ \cmd  \ \ret \x\} : (\overline{\bar{\TW} \to \TW}) \to \bar{\TW} \to \TW}
\end{prooftree}
}
}
\]
\[
\romanobox{
\scalebox{1}{
\begin{prooftree}
\hypo{\pbla  \proc(\bar{\X},\bar{\x})\{ \cmd  \ \ret \x\} : (\overline{\bar{\TW} \to \TW}) \to \bar{\TW} \to \TW}
\hypo{\pbla  \bar{\clos} : \overline{\bar{\TW} \to \TW}}
\hypo{\pbla \bar{\term} : \bar{\TW}}
\infer3[(CALL)]{\pbla \callcc\proc(\bar{\clos},\bar{\term}) : \TW}
\end{prooftree}
}
}
\]

\[
\romanobox{
\begin{prooftree}
\hypo{\typenvs(\va) = \STW}
\infer1[(VAR)]{\pbla \va : \STW}
\end{prooftree}
}
\qquad
\romanobox{
\begin{prooftree}
\hypo{\typenvs\uplus \{\bar{\x} : \bar{\TW}\} ,\typproc,\typop \vdash \term : \TW}
\infer1[(ABS)]{\pbla  \abs{\bar{\x}}.\term : \bar{\TW}\to \TW}
\end{prooftree}
}
\]

\[
\romanobox{
\scalebox{1}{
\begin{prooftree}
\hypo{\pbla \prog : {\STW}}
\hypo{\typproc(\procn) = \langle \typenv, (\sla,\sla_{in},\sla_{out}) \rangle}
\hypo{\pbl \body(\procn) : \sla}
\infer3[(DEC)]{\pbla \letin{{\procn}} \prog : \STW}
\end{prooftree}
}
}
\]

\[
\romanobox{
\begin{prooftree}
\hypo{\typenvs\uplus\{\va:\STW\},\typproc,\typop \vdash \prog : {\STW'}}
\infer1[(BOX)]{\pbla \boite{[\va]}{\prog} : \STW \to \STW'}
\end{prooftree}
}
\]

\hrulefill
\end{minipage}
\caption{Level-based typing rules for second-order programs}
\label{sfig:tsprog}
\end{figure*}

\subsection{Type System}\label{s:type}
Now we adapt the declassification policy presented in Section~\ref{s:declass} to the second-order programming language.

\subsubsection*{Simple Types and Levels.}
Let $\TW$ be the type of words in $\W$. The set $\SST$ contains all \emph{simple types} over $\TW$ that are defined inductively by the following grammar $\STW  ::= \TW \ | \ \STW \to \STW.$

To account for oracle outputs, the considered set of levels is now $\SLI \triangleq \mathbb{N} \cup \{\infty\}$. 
The level $\infty$ behaves as expected: for each $ \sla \in \SLI,\ \sla \ord \infty,\ \sla \mathbf{+} \infty = \infty,\  \sla \join \infty =\infty$, and $\sla \meet \infty =\sla$ hold.
Similarly to simple types, levels can be extended to functional levels by $\STW ::=  \tau \ | \  \STW \to \STW$, with $\tau \in \SLI$.
The order of a type is defined inductively by $\order(\STW_1 \to \STW_2) \triangleq \max(1+\order(\STW_1),\order(\STW_2))$ and $\order(\sla)=\order(\TW) \triangleq 0$.
As usual, an operator $\op$ will have functional levels of the shape $\sla_1 \to \ldots \to \sla_{ar(\op)+1}$ of order $1$ that can thus be viewed as a tuple in $\SLI^{ar(\op)+1}$.

\subsubsection*{Typing Environments.}
We define four kinds of typing environments: 
\begin{itemize}
\item \emph{variable typing environments} $\typenv$ are finite maps in $\Var_0 \to \SLI$,
\item \emph{operator typing environments} $\typop$  map each operator $\op \in \Operator$ to a finite map $\Delta(\op) \in \SL^2 \to\mathcal{P}(\SLI^{ar(\op)+1})$, 
\item \emph{simple typing environments} $\typenvs$ are finite maps in $\Var \to \SST$, 
\item \emph{procedure typing environments} $\typproc$ map each procedure name to a pair $\langle\typenv,\bar{\sla}\rangle$ consisting of a variable typing environment $\typenv$ and a triplet of levels $\bar{\sla} \in \SLI \times \SL \times \SL$.
\end{itemize}
Variable and operator typing environment are defined as in Section~\ref{ss:type}, with the distinction that $\infty$ can appear in the functional level of an operator. Simple typing environments just assign the type $\TW$ to variables in $\Var_0$ and the type $\TW \to \TW$ to variables in $\Var_1$. A procedure typing environment assigns a typing context to each procedure.
Operator, procedure, and simple typing environments are global, i.e. defined for the whole program. Variable typing environments are local, i.e. relative to the procedure under analysis. 

For a procedure typing environment $\typproc$, it will be assumed that for every $\proc \in \dom(\typproc)$, if $\typproc(\proc)=\langle\typenv,\bar{\sla}\rangle$ then $(\param(\proc) \cup \local(\proc)) \cap \Var_0  \subseteq \dom(\typenv)$. I.e., $\typproc$ is defined for the local variables and order-$0$ parameters of the  procedure $\proc$.

\subsubsection*{Typing Judgments.}
The type system uses two kinds of judgments:
\begin{inparaenum}
\item \emph{Statement (resp. expression) typing judgments} $\pbl \cmd : \sla$ (resp. $\pbl \e : \sla$), with $\sla_{in}, \sla_{out} \in \SL$ and $\sla \in \SLI$.  The  meaning of this judgment is that the \emph{statement level} (resp. \emph{expression level}) is $\sla \in \SLI$, under the prerequisite that the \emph{innermost  level} is $\sla_{in}$, and  the \emph{outermost level} is $\sla_{out}$. These levels are defined similarly to Section~\ref{ss:type} and are still finite (i.e., $\neq \infty$) as our type discipline will prevent a loop from being guarded by level $\infty$ data.
\item \emph{Program (resp. term and closure) typing judgments} $\pbla \prog : \STW$ ($\pbla \term :\TW$ and $\pbla \clos : \bar{\TW} \to \TW$, resp.). The meaning of this judgment is that  $\prog$ ($\term$ and $\clos$, resp.) is of simple type $\STW$ ($\TW$ and $\bar{\TW} \to \TW$, resp.) under the considered typing environment. Moreover, if $\order(\STW)=i$ then $\prog$ is said to be of \emph{order-$i$}. On the other hand, a term (resp. closure) is necessarily of order-$0$ (resp. $1$), i.e., of type $\TW$ (resp. $\bar{\TW} \to \TW$). 
\end{inparaenum}

\subsubsection*{Typing Rules.}

The type system is provided in Figures~\ref{sfig:tsprog}.  \emph{Well-typed programs} are normal form and guarded order-$2$ programs that can be given the  type $(\overline{\TW \to \TW}) \to \bar{\TW} \to \TW$, i.e., second-order programs computing a functional. 
The type system is composed of two sub-systems. The typing rules provided at the lower part of Figure~\ref{sfig:tsprog} enforce that terms follow a standard simply-typed discipline.
The typing rules presented in the upper part of Figure~\ref{sfig:tsprog} ensure that procedure bodies follow a level-based type discipline, adapting smoothly the policy of Section~\ref{s:declass} to the second-order setting.
The transition between the two subsystems is performed in the Rule (DEC) of Figure~\ref{sfig:tsprog} that checks that the procedure body follows the level-based type discipline once and for all in a procedure declaration.
In Figure~\ref{sfig:tsprog}, the main differences with  the type system of Figure~\ref{fig:t1ts} are highlighted in \romanotextbox{boxes}.
For simplicity, we have not specified the sequence lengths in the typing rules. However, it is clear that programs must check the following trivial syntactic constraints: in Rule (CALL), the type arity of each closure must match the arity $ar(\X)$ of the corresponding procedure parameter; in Rule (ORC) , the length of $\bar \e$ is equal to $ar(\X)$.

\subsubsection*{Intuition.}We now provide some intuitions on the typing discipline.

The typing rules of Figure~\ref{sfig:tsprog} that are not in \romanotextbox{boxes} implement a standard noninterference policy with declassification, as already presented in Figure~\ref{fig:t1ts}. 
In a loop of level $\sla$, data of level $\sla$ or higher can only be copied or decreased using neutral operators (or declassifications). 
As neutral operators compute subwords or predicates, the corresponding space is polynomially bounded in the size of the program input. The restrictions put on declassified data preserve this property. 
Moreover, data of strictly smaller level can increase by a constant in each assignment. Hence under suitable assumptions: aperiodicity and safety  (see next section), the loop terminates in a polynomial number of steps and the lower level data increase at most polynomially. 
The same kind of reasoning can be performed on loops of strictly smaller level. Hence the program cannot run for time greater than a composition of a constant (in the size of the program) number of polynomials in the input size and the maximal output of an oracle call. This ensure that the program computes a functional in Oracle PolyTime ($\opt$), as discussed in the introduction.

We now discuss the modification on the level-based discipline. Rule (ORC) takes oracle outputs to level $\infty$. 
On this other hand, by the side condition $\tau \neq \infty$, Rules (WH) and (WI) ensure that loops cannot be guarded by data of level $\infty$. 
Consequently, by the side condition $\tau_2 \neq \infty$ of Rule (ASG), oracle outputs cannot be directly assigned to in a loop. Hence the only option for an oracle output to be assigned is to be declassified, i.e., truncated and converted to unary (see the reduction rule for $\declass$ in Figure~\ref{fig:prog2}). We will relax a bit this strong constraint by also allowing the (non unary) truncation of oracle output in the next section.
As a consequence of the restrictions we will put later on admissible operator types and as programs are guarded, only a constant number of unbounded oracle calls can be performed in a typed programs, most of which occur outside while loops (only 1 will be admitted per outermost while loop).
Rule (OBK) enforces that break statements with oracle calls are controlled by expressions of the shape $\size{\X(\bar{\e})} > \size{\X(\bar{\x})}$: the while loop breaks if the oracle call has size greater than the size of the evaluation of $\X(\bar{\x})$. However expressions in $\bar{\x}$ have level $\sla_i$ such that $\sla_{out} \ordst \sla_i$. Hence, they cannot be modified inside the (outermost) while loop. The while loop breaks if the size of the oracle call $\X(\bar{\e})$ exceeds the size of the (constant in the loop) call $\X(\bar{\x})$. Notice that the result of the call does not need to be fully explored for the test to be checked (i.e., if value $\v$ is the result of evaluating $\X(\bar{\x})$ the test can be performed in at most $\size{v}+1$ steps). This means that while we remain in the loop, the oracle calls are all of size bounded by the size of the result of $\X(\bar{\x})$, i.e., the computed functional remains in Finite  Length Revision ($\flr$).

To conclude, the functionals computed by each procedure call are in $\spt = \opt \cap \flr$ and the $\lambda(-)_2$ closure of Theorem~\ref{thm:KS} can just be simulated by our notions of closure and procedure call.

\subsection{Safe Programs and their Properties}\label{ss:sec}

We now adapt the notion of safety to the second-order setting.

\begin{definition}\label{sote}
An operator typing environment $\typop$ is \emph{safe} if $$\Delta(\truncate)(\sla_{in},\sla_{out}) \triangleq \{\infty \to \tau \to \tau'  \ | \  \tau' \ordst \sla_{in} \text{ and }\sla_{out} \ord \tau\}$$ and for each $\op \in \dom(\typop)$, $\op \neq \truncate$, is either neutral, positive, or polynomial, $\sem{\op} \in \FP$, and for each $\sla_{in},\sla_{out} \in \SL$, and for each $\sla_1 \to \ldots \to \sla_{ar(\op)+1} \in \typop(\op)(\sla_{in},\sla_{out})$,
\begin{enumerate}
\item  if $\op$ is a neutral operator then
$\sla_{ar(\op)+1} \ord \meet_{i=1}^{ar(\op)} \sla_i \ord \join_{i=1}^{ar(\op)} \sla_i  \neq \infty$;\label{premier}
\item  if $\op$ is a positive operator then
$\sla_{ar(\op)+1} \ord \meet_{i=1}^{ar(\op)} \sla_i \ord \join_{i=1}^{ar(\op)} \sla_i  \neq \infty,$
and  either $\sla_{ar(\op)+1} \ordst \sla_{in}$ or $\sla_{ar(\op)+1}=\levela$;\label{second}
\item  if $\op$ is a polynomial operator then $\sla_{out}=\levela$.
\label{third} 
\end{enumerate}
\end{definition}

We briefly explain the intuition that lies behind safe operator typing environments. 
Operator $\truncate$ is treated apart to allow non-unary flows from level $\infty$ to a finite level $\tau'$. However the price to pay for that permissiveness is that the output of a truncate can never be used to guard the corresponding loops (as $\tau' \ordst \sla_{in}$).  The other constraints are similar to the ones of Definition~\ref{def:sote}, as there cannot be an operator applied to an oracle output (level $\infty$ data), a declassification or a truncation have to occur first.

\begin{definition}[Safety]
Given a procedure typing environment $\typproc$ and a safe operator typing environment $\typop$, $\prog $ is a \emph{safe program}  if it is a well-typed program wrt. $\Omega$ and $\Delta$, i.e., $ \emptyset,\typproc,\typop \vdash \prog : (\overline{\TW \to \TW}) \to \bar{\TW} \to \TW$ can be derived. Let $\safe$ be the set of safe programs and $\Delta\safe$ be the set of safe programs wrt. $\Delta$.
\end{definition}

\begin{example}\label{ex:safe}
Let us study how procedure K from Listing~\ref{ex:K} can be part of a safe program.
The $\while$ loop can be typed as follows:\\
 
\begin{minipage}{0.42\textwidth}
\begin{lstlisting}[firstnumber=22]
$\while$(s$^{\levelb}$>0)$^{\levelb}${
  $\brk$(|X(i$^{\levela}$)$^{\infty}$|>|X(r$^{\levelc}$)$^{\infty}$|): $\levelb\sep$
  i$^{\levela}\asg\truncate$(X(i$^{\levela}$)$^{\infty}$,p$^{\levelb}$)$^{\levela}: \levela\sep$
  s$^{\levelb}\asg$(s$^{\levelb}$-1)$^{\levelb}$ : $\levelb$
}
 \end{lstlisting}
 \end{minipage}
 
The level of \verb!s! at line 22 is enforced to be at least $\levelb$ (and cannot be $\infty$), by Rule (WI). Hence in the while loop body the innermost and outermost levels are equal to $\levelb$. 
The level of \verb!r! at line 23 is enforced to be $\levelc$, as it needs to be strictly greater than the outermost level, by Rule (OBK). 
The level of \verb!p! at line 24 is enforced to be equal to the outermost level $\levelb$, by definition of safety. 
The level of the whole expression $\truncate$\verb!(X(i),p)! is $\levela$. 
Indeed, by definition of a safe operator typing environment, it has to be smaller than the level of its first operand $\infty$ and strictly smaller than the innermost level $\levelb$. 
Hence at line 24, the level of \verb!i! is enforced to be $\levela$ by Rule (ASG) and the statement can be given the level $\levela$. 
Finally, the full loop body can be typed using the rules for sequence and subtyping (SEQ) and (SUB). 
Notice that the remaining lines of Listing~\ref{ex:K} type as there are no constraints on assignments outside loops.
\end{example}


First, we can show our main result, stating that the set of functionals computed by safe, terminating, and aperiodic programs is exactly $\BFF$. Hence, we have succeeded to capture this class using the $\spt$ scheme of Theorem~\ref{thm:KS}.

\begin{restatable}{theorem}{thmmain}\label{thm:main}
$\sem{\safe  \cap \ASN} = \BFF$.
\end{restatable}

And we can show that safety is tractable. Again, let $\size{\prog}$ be the number of symbols in $\prog$.
\begin{restatable}{theorem}{thmtiprog}\label{thm:tiprog}
Given a safe operator typing environment $\typop$, deciding whether $\prog \in \Delta\safe$  can be done in time polynomial in $\size{\prog}$.
\end{restatable}

\section{Conclusion and Future Work}\label{s:con}
We have provided a new declassification policy for program complexity analysis and shown that it can be applied to provide an expressive characterization of $\FP$ and the first characterization of the class $\BFF$ based on $\spt$. 
The declassification policy uses a declassification construct that can be used  inside while loop and is, hence, quite general. The characterization is achieved by putting extra and intuitive restrictions: aperiodicity and termination in the second-order case.  We also show that type inference for safety is tractable and that aperiodicity is a $\Pi^0_1$-complete problem. We have also exhibited a tractable and completeness-preserving criterion for aperiodicity. A future line of research is the search for tractable and more expressive criteria for aperiodicity.

\bibliography{bib}
\end{document}
\newpage

\appendix

\section{Proof of Type-1 Completeness}
\begin{example}\label{ex:mult1}
Consider the program below computing the multiplication.
 
\begin{lstlisting}[linewidth=.42\textwidth]
mult(x$^\levelc$,y$^\levelc$) {
  r$^\levelb\asg$0 :$\levelc\sep$
  $\while$(y$^\levelc$>0){
    z$^\levelc\asg$x$^\levelc$ :$\levelc\sep$
    $\while$(z$^\levelc$>0){
      r$^\levelb\asg$r+1 :$\levelc\sep$
      z$^\levelc\asg$z-1  :$\levelc$
    };
    y$^\levelc\asg$y$^\levelc$-1 :$\levelc$
  }
  $\ret$r
}
\end{lstlisting}

levels of variables ($\typenv$) have been provided as superscripts in the code.
The following levels are included in $\typop$:

$$
\begin{array}{rcl}
\levelc\to\levelc\to\levelc&\in&\typop(\texttt{and})(\levelc, \levelc)\cap\typop(>)(\levelc,\levelc)\\
\levelb\to\levelb & \in & \typop(+1)(\levelc, \levelc)\\
\levelc\to\levelc & \in & \typop(-1)(\levelc, \levelc).
\end{array}
$$
This typing environment is safe as \texttt{and}, $>$, and $-1$ are neutral operators and $+1$ is a positive operator.


The program is aperiodic since, for the enclosing while loop (lines 3 to 10), $\store(\y)$ is different each time the guard $\y>0$ is encountered (it has been decremented).
Similarly, for the inner loop (lines 5 to 8), only unique values $\store(\z)$ are seen in the strict subtrees of root $(\while(\z>0)\{\ldots\},\store)$.
\end{example}

\thmfpcomplete*

\begin{proof}
Given function $f$ computed by a Turing machine $M_f$ in time $Q(n)$ for inputs of size $n$, we encode this computation by first computing the time bound, second defining a statement that simulates the one-step transition of the Turing machine, and finally iterating this statement $Q(n)$ times. 
The computation of polynomials is a matter of multiplications and additions.
We know, from Example~\ref{ex:mult1}, how to do it safely and aperiodically, and obtain the result in a variable \texttt{r} of level $\levelb$.

We can encode the configuration of the Turing machine using 3 variables of level $\levela$ that represent the left part of the tape, the right part of the tape, and the state of the machine.
The operations on those variables are not neutral but positive (with constant 1 for the operators on tapes and constant the number of states for the state variable).
 The transition function of $M_f$ is then implemented by a $\instr{if}$ statement that reads and modifies those variables. This statement can be given level $\levela$.
Assume the initial state of the machine is $0$, we obtain a program of the following form:

\begin{lstlisting}[linewidth=0.42\textwidth]
prog$_f$(x$^\levelc$){
    r$^{\levelb}\asg$0$\sep$
    while(...$^\levelc$){
        ...
        //compute r$^{\levelb}$ equal to $Q(|\store($x$)|)$
    }$\sep$
    left$^{\levela}\asg\epsilon$$\sep$
    right$^{\levela}\asg$x$^\levelc$$\sep$
    q$^{\levela}\asg$0$\sep$
    while(r$^{\levelb}$>0){
        r$^\levelb\asg$r-1$\sep$
        h$^{\levela}\asg$head(right$^{\levela}$)$\sep$
        /*List of ifs simulating 
          the transition function */
          if(h$^{\levela}$=$\alpha$ and q$^{\levela}$=10) {
            /*$\text{E.g., when the symbol is }\alpha$
            $\text{and the state is 10,}\ \text{write }\beta\text{,}$
            $\text{move right, enter state 11}$*/
            left$^{\levela}\asg$cons($\beta$,left$^{\levela}$)$\sep$
            right$^{\levela}\asg$tail(right)$\sep$
            q$^{\levela}\asg$11
        }
        ...
    }
    return right$^{\levela}$
}
\end{lstlisting}

The variable typing environment $\typenv$ is provided by the variable annotations in the program code. 
The safe operator typing environment $\typop$ is an extension of the one from Example~\ref{ex:mult1} with
$$\arraycolsep=3pt
\begin{array}{rcl}
\{\levelb\to\levelb\} & \subset & \typop(>0)(\levelb, \levelb)\\
\{\levelb\to\levelb\} & \subset & \typop(-1)(\levelb, \levelb)\\
\forall\sla_1, \sla_2, \{\levela\to\levela\to\levela\} & \subset & \typop(=)(\sla_1, \sla_2) \cap \typop(\texttt{and})(\sla_1, \sla_2)\\
\forall\sla_1,\sla_2, \{\levela\to\levela\} & \subset & \typop(+1)(\sla_1, \sla_2)\cap\typop(\texttt{head})(\sla_1, \sla_2)\\
&&\cap\typop(\texttt{cons})(\sla_1, \sla_2)\cap\typop(\texttt{tail})(\sla_1, \sla_2)\\
\end{array}
$$

This program is safe and aperiodic.
\end{proof}

\section{Proof of Type-1 Soundness}
Given a typing judgment $j$, a \emph{typing derivation} $\rho \rightslice j$  is a tree whose root is the typing judgment $j$ and whose children are obtained by applications of the typing rules of Figure~\ref{fig:t1ts}. The notation $\rho$ will be used whenever mentioning the root of a typing derivation is not explicitly needed. 
Given two typing derivations $\rho$ and $\rho'$, we write $\rho' \sqsubseteq \rho$ (respectively $\rho' \sqsubset \rho$) if $\rho'$ is a (strict) subtree of $\rho$. $\rho'$ is called a \emph{subderivation} of $\rho$. 

To prove soundness, we introduce intermediate lemmas.
The following Lemma states that statement levels are monotonic in their substatement level in a given typing derivation.

\begin{lemma}[Monotonicity]\label{lem:monotonicity}
For any two typing subderivations $\rho \rightslice \pbl \cmd : \sla$ and $ \rho' \rightslice \pblp \cmd' : \sla'$ of a safe program, if $\rho' \sqsubseteq \rho$ then $\sla' \ord \sla$. 
\end{lemma}

\begin{proof}
Suppose by contradiction that $\rho' \sqsubseteq \rho$ and $\sla \ordst \sla'$ hold. As all typing rules for statements in Figure~\ref{fig:t1ts} 
are monotonic in the statement level,  $\rho \rightslice \pbl \cmd : \sla$ cannot be derived from $\rho'$.\qedhere
\end{proof}

Let $\A(\cmd)$ be the set of variables that are assigned to in $\cmd$, for example, $\A(\x \asg \y \sep \y \asg \z)= \{\x,\y\}$.
The confinement Lemma expresses that statements of level $\sla$ cannot write in variables of strictly higher level.

\begin{lemma}[Confinement]\label{lem:confinement}
For any typing subderivation $\rho \rightslice \ \pbl \cmd:\sla$ of a safe program, it holds that for all $ \x \in \A(\cmd)$, $\Gamma(\x) \ord \sla$.
\end{lemma}  

\begin{proof}
By contradiction. Suppose that $\rho \rightslice \pbl \cmd:\sla$ holds. Consider a variable $\x \in \A(\cmd)$ and suppose that $\sla \ordst \Gamma(\x)$. By typing Rule (ASG), there is an expression $\e$  and there are levels $\sla_{in}',\sla_{out}'$ such that $\rho' \rightslice  \vdash_{\sla_{out}'}^{\sla_{in}'} \x \asg \e : \Gamma(\x)$ and $\rho' \sqsubseteq \rho$. 
Consequently, by Lemma~\ref{lem:monotonicity}, $ \Gamma(\x) \ord \sla$, which contradicts the assumption.
\qedhere
\end{proof}

%
%
%

Let $\diamond$ be a special symbol denoting a hole. Given a sequence of expressions $\bar{\e}^n$ of length $n$, let $\bar{\e}_{i,C}^n$, for $i \leq n$, denote the sequence obtained by replacing $\e_i$ by $C$ in $\bar{\e}$. Contexts are defined by the following grammar:
\[
C, D, \ldots 
   \rgl \diamond
   \ |\ \op(\bar{\e}_{i,C}^{ar(\op)})  \ | \ 
   \declass(\bar{\e}_{i,C}^2)\ | \ 
   \bar{\e}_{i,C}^n
\]

Given an expression $\e$ and a context $C$, $C[\e]$ is the expression obtained as the result of substituting  the expression $\e$ for the hole $\diamond$.
In this setting, $\e$ is a \emph{subexpression} of $C[\e]$.
Given two expressions $\e$ and $\e'$, $\e'$ is \emph{declassified in }$\e$ if there are two contexts $C,D$ and an expression $\d$ such that $\e=C[\declass(D[\e'],\d)]$.

In a safe program, only variables of higher level or whose occurrences are all declassified 
can be accessed when evaluating an expression.

\begin{lemma}\label{lem:ss}
For any triplet of typing subderivations $$
\begin{array}{l}
\rho \rightslice \pbl  \while (\e) \{\cmd\} : \sla, \\
\rho' \rightslice  \pblp \e' : \sla',\text{ and}\\
\rho'' \rightslice \pblpp \e'' : \sla''
\end{array}
$$ of a safe program, if $\rho'' \sqsubseteq \rho' \sqsubseteq \rho$, then (at least) one of the following conditions holds:
\begin{enumerate}
\item $\sla' \ord \sla''$,
\item $\e''$ is declassified in $\e'$.
\end{enumerate}
\end{lemma}

\begin{proof}
 By structural induction on the expression $\e'$. 
 \begin{itemize}
\item Case $\e'=\x=\e''$: $\sla' = \sla''$. 
 
\item Case $\e'=\op(\ldots,\e^3,\ldots)$: Assume that $\e^3$ is of level $\sla^3$. If $\e''$ is declassified 
in $\e^3$ then $\e''$ is declassified 
in $\e'$; if $\sla^3 \ord \sla''$ then $\sla' \ord \sla^3 \ord \sla''$, by definition of safe operator typing environment for neutral and positive operators ($\op$ cannot be polynomial in a loop) and by applying typing Rule (OP).

\item Case $\e'=\declass(\e^3,\d)$: if $\e''\in \e^3$ then $\e''$ is declassified in $\e'$. If $\e'' \in \d$ of level $\sla^3$ then we apply the induction hypothesis: if $\e''$ is declassified 
in $\d$ then it is declassified 
in $\e'$; if $\sla^3 \ord \sla''$ then, by typing Rule (DCL), $\sla^3$ is equal to the outermost level $\sla_{out}'$ and then $\sla' \ord \sla^3 \ord \sla''$.\qedhere
 \end{itemize}

 \end{proof}

Consequently, the innermost level of a while loop provides a lower bound on the level of the variables in the guard that are not declassified.

The following Lemma shows that the innermost level of a while loop substatement is always an upper bound on the level of this loop. 

\begin{lemma}\label{lem:inner}
For any pair of typing subderivations  $ \rho \rightslice \pbl \while (\e) \{\cmd\}   : \sla$ and $\rho' \rightslice \pblp \cmd' : \sla'$ of a safe program, if $ \rho' \sqsubseteq \rho$, then $\sla'_{in} \ord \sla$.
\end{lemma}
\begin{proof}
By a case analysis on rules (WI) and (WH) and using Lemma~\ref{lem:monotonicity}.
\qedhere
\end{proof}

We now show that within a while loop of level $\sla$, if an expression involving positive operators is assigned to a level $\sla$ variable then the positive subexpressions are all declassified.

\begin{lemma}[Stratification]\label{lem:stratification1}
For any typing subderivations $\rho \rightslice \pbl  \while (\e) \{\cmd\} : \sla$ and $ \rho' \rightslice  \pblp \x' \asg \e' : \sla'$ of a safe program such that $\rho' \sqsubset \rho$, if $\sla =\sla'$ then for each $\e'' \in \e'$,
if $\e''=\op(\bar{\e})$, for some positive operator $\op$,
then $\e''$ is declassified in $\e'$.
\end{lemma}
\begin{proof}
The proof is by induction on the size of the expression $\e'$.
Consider a subexpression $\e'' \in \e'$. 
Let $\e''=\op(\bar{\e})$, for some positive operator $\op$, then by definition of safe operator typing environment and by Rule (OP), it holds that if $\op(\bar{\e})$ is of level $\sla''$ then $\sla'' \ordst \sla'_{in}$. Consequently, $\sla'' < \sla$, by Lemma~\ref{lem:inner}. As the level of $\e'$ is greater than $\sla$ by Rule (ASG) then $\op(\bar{\e})$ is declassified in $\e'$, by Lemma~\ref{lem:ss}.
\qedhere
\end{proof}

Now we show that in safe programs any assignment in a loop can cause the memory size to increase by at most a constant.

\begin{lemma}[Controlled increase]\label{lem:ci}
For any typing subderivations $\rho \rightslice \pbl  \while (\e) \{\cmd\} : \sla$ and $ \rho' \rightslice  \pblp \x' \asg \e' : \sla'$ of a safe program such that $\rho' \sqsubset \rho$, $\exists c \in \mathbb{N}$ such that $\forall \store \in {\tt Stores}$, if $(\store,\e') \toe \w$ then $\size{\w} \leq \max_{\x \in \dom(\store)}(\size{\store(\x)})+c.$
\end{lemma}
\begin{proof}
This proof is by induction on expressions. 
First note that as the program is safe, by definition of safe operator typing environment (Definition~\ref{sote}), no polynomial operator can be called in $\e'$, as the corresponding assignment is inside a while loop by hypothesis ($\rho' \sqsubset \rho$).
Hence its outermost level $\sla_{out}'$ is distinct from $\levela$.
\begin{itemize}
\item Case $\e' = \y$. Then by reduction Rule (Var), $\w=\store(\y) \leq \max_{\x \in \dom(\store)}(\size{\store(\x)})$.
\item Case $\e' = \op(\bar{\e})$.
By induction hypothesis, if $(\store,\e_i) \toe \w_i$ then $$\size{\w_i} \leq \max_{\x \in \dom(\store)}(\size{\store(\x)})+c_i.$$ By definition of neutral and positive operators, it holds that $\size{\w}=\size{\sem{\op}(\bar{\w})} \leq \max_i(\size{\w_i})+d$, for some $d \geq 0$. Hence $\size{\w} \leq \max_{\x \in \dom(\store)}(\size{\store(\x)})+ \max_i c_i +d$. Take $c \triangleq \max_i c_i +d$. 
\item Case $\e' = \declass(\e_1,\e_2)$. Rule (DCL) implies that the level of $\e_2$ is $\sla_{out}$, which in the body of a while loop needs to be finite, from (WI) and (WH). We can apply the induction hypothesis on $\e_2$: assume $(\store,\e_2) \toe \w_2$ then $\size{\w_2} \leq \max_{\x \in \dom(\store)}(\size{\store(\x)})+c_2$. If $(\store,\e') \toe \w$, then from the semantics of $\declass$, it holds that:
$$
\size{{\w}} \leq \size{\w_2} \leq \max_{\x \in \dom(\store)}(\size{\store(\x)})+c_2.
$$\qedhere
\end{itemize}
\end{proof}

Now we show that in a loop of level $\sla$, expressions assigned to variables of level greater than $\sla$ and also the loop guard are restricted to some particular bounded domain. For that purpose, we define a notion of span,  corresponding to the set of values that are reachable from a given store.

\begin{definition}[Span]
Given a store $\store$, a typing environment $\typenv$ such that $\dom(\store)=\dom(\typenv)$, and a level $\sla$, the \emph{span} of $\store$ wrt $\typenv$ and $\sla$ is defined by
\[
\begin{array}{rcl}
\Span& \triangleq &\{ \w \ | \ \exists \x \in \dom(\store),\ \w \dord \store(\x) \wedge \sla \ord \typenv(\x)\} \cup\\
&& \{ 1^n \ | \ n \leq \max_{\x \in \dom(\store) \wedge \sla \ord \typenv(\x)}\size{\store(\x)}\}\cup \{0,1\}.
\end{array}
\]
\end{definition}
Notice that $\Span$ is closed under subword, i.e., if $\w \in \Span$ and $\w' \dord \w$ then $\w' \in \Span$.

\begin{lemma}[Bounded span]\label{lem:bg}
For any typing subderivations $\rho \rightslice \pbl  \while (\e) \{\cmd\} : \sla$ and $ \rho' \rightslice  \pblp \x' \asg \e' : \sla$ of a safe program such that $\rho' \sqsubset \rho$ holds, $\forall \store \in \texttt{Stores}$, if $(\store,\e') \toe \w$ then $\w \in \Span$.
\end{lemma}
\begin{proof}
By induction on expressions. 
First note that as the program is safe, by definition of safe operator typing environment (Definition~\ref{sote}), no polynomial operator can be called in $\e'$, as the corresponding assignment is inside a while loop by hypothesis ($\rho'  \sqsubset \rho$). Hence its outermost level $\sla_{out}'$ is distinct from $\levela$.
\begin{itemize}
\item Case $\e' = \y$. Then by reduction Rule (Var), $\w=\store(\y)$. Moreover, by typing Rule (VAR), $\sla = \typenv(\y)$. Hence $\w \in \Span$.
\item Case $\e' = \op(\bar{\e})$. $\op$ cannot be positive, by Lemma~\ref{lem:stratification1}. If $\op$ is a Boolean predicate then $\w \in \Span$ trivially holds (as $0,1 \in \Span$ holds). If $\op$ is not a Boolean predicate then, assume by induction hypothesis, that for each $i \leq \ell(\bar{\e})$, if $(\store,\e_i) \toe \w_i$ then $\w_i \in \Span$. It trivially holds that $\sem{\op}(\bar{\w}) \in \Span$, as $\Span$ is closed by subword. 
\item Case $\e' = \declass(\e_1,\e_2)$ with $(\store,\e_i) \toe \w_i$. According to the semantics of $\declass$, $\w=1^n$, for some $n \leq \size{\w_2}$. Moreover, by typing Rule (DCL), $\e_2$ is of level $\sla'_{out}$ and it holds that $\sla \ord \sla'_{out}$. Hence by induction hypothesis, $\w_2 \in \Span$. It also implies that $\w = 1^n\in \Span$.\qedhere
\end{itemize}
\end{proof}

Given a variable typing environment $\typenv$ and a level $\sla$, we define an order relation over stores by $\store \lesssim_{\sla}^\typenv \store'$ iff $\Span \subseteq \Spanp$.

\begin{lemma}[Stability]\label{lem:stab}
For any typing subderivation $\rho \rightslice \pbl  \while (\e) \{\cmd\}: \sla$ of a safe program, for any $\store,\store'$, and $\cmd'$ such that $\pi_{(\store', \cmd')}$ is a subtree of $\pi_{(\store, \while (\e) \{\cmd\})}$, we have $\store' \lesssim_{\sla}^{\typenv} \store$.
\end{lemma}
\begin{proof}
By induction on the evaluation tree.
In the evaluation tree, the store $\store$ is updated by assignments of the form $\x \asg \e'$. 
If $\typenv(\x)\ordst\sla$ and $(\store'', \e') \toe \w$, for some $\store''$ then $\Spanp = \Spanpp \subseteq \Span$, by induction hypothesis.
Otherwise $\typenv(\x) = \sla$, by Lemma~\ref{lem:confinement}.  
If $(\store'', \e') \toe \w$, for some $\store''$ then $\w\in\Span$, by Lemma~\ref{lem:bg}. Moreover, by induction hypothesis, $\store'' \lesssim_{\sla}^\typenv \store$.  Hence $\store'(\x) \in \Spanpp$ and $\store' \lesssim_{\sla}^\typenv \store''\lesssim_{\sla}^\typenv \store$. \qedhere
\end{proof}

As values in the span are linearly bounded by the size of the input store (and some constant for boolean values), we obtain the following corollary.
\begin{corollary}\label{coro:bound}
For any pair of typing derivations $\rho \rightslice \pbl  \while (\e) \{\cmd\}: \sla$ and $\rho' \rightslice \pblp \e' : \sla'$ of a safe program such that $\rho' \sqsubseteq \rho$ and $\sla \ord \sla'$, for any $\store,\store',\sigma,\phi,$ such that $\pi_{(\sigma, \store', \phi, \e')} \sqsubseteq \pi_{(\sigma, \store, \phi, \while (\e) \{\cmd\})}$, if $(\sigma, \store', \phi, \e') \toexp \w$ then $$\size{\w} \leq \max(1,\max_{\y \in dom(\store) } \{\size{\store(\y)} \ | \  \sla \ord \typenv(\y)\}).$$
\end{corollary}

\thmpoli*
\begin{proof}
Let $\prog$ be a safe aperiodic terminating program. Consider an outermost while loop: 
$\rho \rightslice \typenv,\typop \vdash_{\levela}^{\levela} \while(\e)\{\cmd\}:\sla$ in $\prog$.
For any subexpression $\e'$ of $\e$, from Lemma~\ref{lem:ss}, either the level $\sla$ of $\e$ is less or equal to that of $\e'$, or $\e'$ is declassified in $\e$.
In this second case, we can write $\e=C[\declass(\e_1, \e_2)]$ for some context $C$ containing no declassification.
It implies that all expressions in $C$ have level greater or equal to $\sla$.
By Lemma~\ref{lem:stab}, for any expression $\e'$ in $C$ and any store $\store'$ in the execution tree $\pi_{(\store, \while(\e)\{\cmd\})}$, if $(\store', \e') \toe \w$, then $\w\in\Span$
By Corollary~\ref{coro:bound}, $(\store', \declass(\e_1, \e_2)) \toe^* 1^n\in\Span$.
There exists a constant $c$, bounded by the size of $\e$, such that
the number of distinct values that $\e$ can reach is bounded by $(\#\Span)^c$.
The aperiodicity hypothesis hence implies that the guard is evaluated at most a number of time polynomial in the size of $\store$.
Using Lemma~\ref{lem:ci}, the values of level less than $\sla$ can increase at most polynomially in $\max_{\x}(\size{\store(\x)})$ (apply a constant increment a polynomial number of times).
Hence, we can apply the same reasoning on inner loops.
The case of sequential outer loops relies on the fact that between loops, it is possible to apply polynomial operators a constant number of times.
Hence the complexity of the program is bounded by a composition of polynomials, which is a polynomial.
\end{proof}

\section{Type-2 Soundness and Completeness}

For the higher-order language of section~\ref{s:prog}, some new lemmas are needed and some previous lemmas need modifications.
The monotonicity lemma (Lemma~\ref{lem:monotonicity}) and the confinement lemma (Lemma~\ref{lem:confinement}) have already been proven.

\begin{lemma}\label{lem:nonboundediterationinloops}
Let $\rho \rightslice \pbl \while (\e) \{\cmd\} : \sla $ be a typing subderivation of a safe program. For any $ \x \in \A(\cmd)$, $\Gamma(\x) \neq \infty$.
\end{lemma}
\begin{proof}
By the side conditions of Rules (WH) and (WI). It holds that $\sla \neq \infty$. Consequently, the result holds by applying Lemma~\ref{lem:confinement}.
\qedhere
\end{proof}

In Lemma~\ref{lem:ss}, a third case can arise when oracles appear.
Contexts are defined by the following grammar:
\begin{align*}
C, D, \ldots 
	\rgl &\diamond
    \ |\ \op(\bar{\e}_{i,C}^{ar(\op)})  \ | \truncate(\bar{\e}_{i,C}^2) \ | \\
    &  \declass(\bar{\e}_{i,C}^2)\ | \ \X (\bar{\e}_{i,C}^n) \ | \ \bar{\e}_{i,C}^n
\end{align*}

Given two expressions $\e$ and $\e'$, $\e'$ is \emph{oraculated in }$\e$ if $\e=C[\X(D[\e'])]$.

In a safe program, only variables of higher level or whose occurrences are all declassified or oraculated
can be accessed when evaluating an expression.

\begin{lemma}\label{lem:ss2}
For any triplet of typing subderivations $\rho \rightslice \pbl  \while (\e) \{\cmd\} : \sla $, $ \rho' \rightslice  \pblp \e' : (\sla',\slb')$, and $\rho'' \rightslice \pblpp \e'' : (\sla'',\slb'')$ of a safe program, if $\rho'' \sqsubseteq \rho' \sqsubseteq \rho$, then (at least) one of the following conditions holds:
\begin{enumerate}
\item $\sla' \ord \sla''$,
\item $\e''$ is declassified in $\e'$,
\item $\e''$ is oraculated in $\e'$.
\end{enumerate}
Moreover, if $\e'$ is the guard of a while loop then Condition 3 cannot hold.
\end{lemma}

\begin{proof}
In the structural induction on the expression $\e'$, the cases already proved  stay true (but occurences of ``declassified'' are to be understood as ``declassified or oraculated'').
$\truncate$ and $\X(.)$ need to be explored:
\begin{itemize}
\item Case $\e'=\truncate(\e^3,\d)$: is similar to the case of operators. Indeed, assume that $\sla^3$ is the level of $\e^3$. By safety, the level of $\d$ is enforced to be $\sla_{out}'(=\sla_{out}'')$ and it always holds that $\sla' \leq  \sla^3 \meet \sla_{out}''$. Hence we can apply the induction hypothesis as in the previous case.
\item Case $\e'=\X(\bar{\d})$: if $\e'' \in \bar{\d}$ then $\e''$ is oraculated in $\e'$.
\end{itemize}

Now, we show that if $\e'$ is the guard of a while loop then either Condition 1 or Condition 2 hold. For that purpose, we show that if $\e''$ is oraculated then $\e''$ is also declassified. First notice that, by typing rules (WH) and (WI), the level of $\e'$ is distinct from $\infty$. Assume that $\e''$ is oraculated in $\e'$. This means that there is $\e^3 = \X(\bar{\d})$ of level $\infty$ such that  $\e'' \in \e^3$ and $\e^3 \in \e'$ both hold. However, by definition of safe operator typing environments and by typing Rule (ORC), an operator and an oracle call cannot be applied to $\e^3$. A declassification cannot be applied as, by Rule (DCL), the level has to be smaller than the outermost level, which is finite. Hence for $\e'$ to be typed, the subexpression $\X(\d)$ has to be truncated. However, by safety, the level of an expression $\truncate( \X(\d), \d')$ is strictly smaller than the innermost level. Hence for $\e'$ to type, some declassification has to be applied (at least one). Consequently, $\e''$ is declassified in $\e'$.
\end{proof}

\begin{lemma}[Stratification II]\label{lem:stratification2}
For any typing subderivations $\rho \rightslice \pbl  \while (\e) \{\cmd\} : \sla$ and $ \rho' \rightslice  \pblp \x' \asg \e' : \sla'$ of a safe program such that $\rho' \sqsubset \rho$, if $\sla =\sla'$ then for each $\e'' \in \e'$, if one of the following condition holds:
\begin{itemize}
\item $\e'' = \X(\bar{\e})$,
\item $\e'' = \truncate(\d,\d')$,
\item $\e''=\op(\bar{\e})$, for some positive operator $\op$,
\end{itemize}
then $\e''$ is declassified in $\e'$.
\end{lemma}
\begin{proof}
The proof is by induction on the size of the expression $\e'$.
Consider a subexpression $\e'' \in \e'$. 

i) If $\e'' = \X(\bar{\e})$, then it has level $\infty$. As such a level cannot  be assigned to in a while loop, by Lemma~\ref{lem:nonboundediterationinloops}, then $\e''$ has to be truncated for the program to type. Just note in the typing rules that $\truncate$ is the only operator allowing a flow from $\infty$ to some finite level. Hence it suffices treat case ii) under the assumption that $\e''$ cannot be oraculated, by side conditions of Rule (ORC) which prevents oracle composition.

ii) If $\e''=\truncate(\d,\d')$, then by safety, it holds that if $\e''$ is of level $\sla''$ then $\sla'' \ordst \sla'_{in}$. Consequently, $\sla'' < \sla$, by Lemma~\ref{lem:inner}. As the level of $\e'$ is greater than $\sla$ by Rule (ASG) then $\e''$ is either declassified or oraculated in $\e'$, by Lemma~\ref{lem:ss2}. The case where $\e''$ is oraculated can be treated using i).

iii) If $\e''=\op(\bar{\e})$, for some positive operator $\op$, then by definition of safe operator typing environment and by Rule (OP), it holds that if $\op(\bar{\e})$ is of level $\sla''$ then $\sla'' \ordst \sla'_{in}$. Consequently, $\sla'' < \sla$, by Lemma~\ref{lem:inner}. As the level of $\e'$ is greater than $\sla$ by Rule (ASG) then $\op(\bar{\e})$ is either declassified or oraculated in $\e'$, by Lemma~\ref{lem:ss2}. The case where $\e''$ is oraculated can be treated using i).
\qedhere
\end{proof}

\begin{lemma}[Controlled increase II]\label{lem:ci2}
For any typing subderivations $\rho \rightslice \pbl  \while (\e) \{\cmd\} : \sla$ and $ \rho' \rightslice  \pblp \x' \asg \e' : \sla'$ of a safe program such that $\rho' \sqsubset \rho$, $\exists c \in \mathbb{N}$ such that $\forall (\sigma,\store,\phi) \in  ({\tt Procedures}) \times {\tt Stores} \times {\tt Env} $, if $(\sigma,\store,\phi,\e') \toexp \w$ then $$\size{\w} \leq \max_{\x \in \dom(\store)}(\size{\store(\x)})+c.$$
\end{lemma}
\begin{proof}
From Lemma~\ref{lem:nonboundediterationinloops}, $\typenv(\x)\neq \infty$ and the level of $\e'$ is finite.
This proof is by induction on expressions whose level is not $\infty$.
As the program is safe, by definition of safe operator typing environment (Definition~\ref{sote}), no polynomial operator can be called in $\e'$, as the corresponding assignment is inside a while loop by hypothesis ($\rho' \sqsubset \rho$). Hence its outermost level $\sla_{out}'$ is distinct from $\levela$.
\begin{itemize}
\item Case $\e' = \y$. Treated in Lemma~\ref{lem:ci}
\item Case $\e' = \op(\bar{\e})$. By rule (OP), $\forall i\leq \ell(\bar{\e})$, $\e_i$ is distinct from $\infty$, so it is as in Lemma~\ref{lem:ci}
\item Case $\e' = \declass(\e_1,\e_2)$ has been treated in Lemma~\ref{lem:ci}
\item Case $\e' = \truncate(\e_1,\e_2)$ is similar to $\declass$.
Assume $(\sigma,\store,\phi,\e_2) \toexp \w_2$ then $\size{\w_2} \leq \max_{\x \in \dom(\store)}(\size{\store(\x)})+c_2$.
If $(\sigma,\store,\phi,\e') \toexp \w$ then by Rule (Truncate), it holds that:
$$
\size{{\w}} \leq \size{\w_2} \leq \max_{\x \in \dom(\store)}(\size{\store(\x)})+c_2.
$$\qedhere
\end{itemize}
\end{proof}

\begin{lemma}[Bounded span II]\label{lem:bg2}
For any typing subderivations $\rho \rightslice \pbl  \while (\e) \{\cmd\} : \sla$ and $ \rho' \rightslice  \pblp \x' \asg \e' : \sla$ of a safe program such that $\rho' \sqsubset \rho$ holds, $\forall (\sigma,\store,\phi) \in \mathcal{P}({\tt Procedures}) \times {\tt Stores} \times {\tt Env} $, if $(\sigma,\store,\phi,\e') \toexp \w$ then $\w \in \Span$. 
\end{lemma}
\begin{proof}
The proof is the same as that of Lemma~\ref{lem:bg}, but
by induction on expressions of level distinct from $\infty$.
Cases $\e' = \truncate(\e_1,\e_2)$ and $\e' = \X(\bar{\e})$ cannot occur, by Lemma~\ref{lem:stratification2}. 
\end{proof}
\begin{lemma}[Stability II]\label{lem:stab2}
For any typing subderivation of a safe program $\rho \rightslice \pbl  \while (\e) \{\cmd\}: \sla$, for any $\store,\store',\sigma,\phi,$ and $\cmd'$ such that $\pi_{(\sigma, \store', \phi, \cmd')}$ is a subtree of $\pi_{(\sigma, \store, \phi, \while (\e) \{\cmd\})}$, we have $\store' \lesssim_{\sla}^{\typenv} \store$.
\end{lemma}
\begin{proof}
By induction on the evaluation tree.
In the evaluation tree, the store $\store$ is updated by assignments of the form $\x \asg \e'$. 
If $\typenv(\x)\ordst\sla$ and $(\sigma, \store'', \phi, \e') \toexp \w$, for some $\store''$ then $\Spanp = \Spanpp \subseteq \Span$, by induction hypothesis.
Otherwise $\typenv(\x) = \sla$, by Lemma~\ref{lem:confinement}.  
If $(\sigma, \store'', \phi, \e') \toexp \w$, for some $\store''$ then $\w\in\Span$, by Lemma~\ref{lem:bg2}. Moreover, by induction hypothesis, $\store'' \lesssim_{\sla}^\typenv \store$.  Hence $\store'(\x) \in \Spanpp$ and $\store' \lesssim_{\sla}^\typenv \store''\lesssim_{\sla}^\typenv \store$. \qedhere
\end{proof}
\begin{corollary}\label{coro:bound2}
For any typing derivations $\rho \rightslice \pbl  \while (\e) \{\cmd\}: \sla,\ \rho' \rightslice \pblp \e' : \sla'$ of a safe program such that $\rho' \sqsubseteq \rho$ and $\sla \ord \sla'$, for any $\store,\store',\sigma,\phi,$ such that $\pi_{(\sigma, \store', \phi, \e')} \sqsubseteq \pi_{(\sigma, \store, \phi, \while (\e) \{\cmd\})}$, if $(\sigma, \store', \phi, \e') \toexp \w$ then $$\size{\w} \leq \max(1,\max_{\y \in dom(\store) } \{\size{\store(\y)} \ | \  \sla \ord \typenv(\y)\}).$$
\end{corollary}
\subsubsection*{Polynomial Step Count.}\label{ss:psc}
In this section, we show that safe and aperiodic terminating programs have a runtime polynomially bounded by the size of the input store and the maximal size of answers returned by the oracle in the course of evaluation.

\begin{definition}
Given a program $\prog$, an input store $\store\triangleq \store_\emptyset[\bar{\x} \leftarrow \bar{\w},\bar{\X} \leftarrow \bar{f}]$, and a procedure $\proc$ of $\prog$,
let $m_{\store}^{\prog,\proc}$ be the maximum of $\size{\store}$ and the maximum size of an oracle answer in any subderivation $\pi_{(\sigma,\store',\callcc\proc(\bar{\clos},\bar{\term}))}$ of $\pi_{(\emptyset,\store,\prog)}$.
    Formally, \begin{align*}
    or_{\store}^{\prog,\proc} &\triangleq \max_{\scalebox{0.8}{$\pi_{(\sigma,\store',\callcc\proc(\bar{\clos},\bar{\term}))} \sqsubseteq \pi_{(\emptyset,\store,\prog})$ }}\{\size{\w};\w\in C(\pi_{(\sigma,\store',\callcc\proc(\bar{\clos},\bar{\term}))})\} \\
 &\text{where }C(\pi_{(\sigma,\store',\callcc\proc(\bar{\clos},\bar{\term}))})\text{ is the set of }\w\text{ such that:}\\
   & \scalebox{1}{\begin{prooftree}
\hypo{\cdots}
\infer1[(Orc)]{(\sigma,\store,\phi \Imp \X(\bar{\e})) \toexp \w}
\end{prooftree}}
\in \pi_{(\sigma,\store',\callcc\proc(\bar{\clos},\bar{\term}))}\\
    m_{\store}^{\prog,\proc} &\triangleq \max(\size{\store}, or_{\store}^{\prog,\proc})
    \end{align*}

A procedure $\proc$ of the program $\prog$ has \emph{Polynomial Step Count} if there is a polynomial $Q \in \mathbb{N}[X]$ such that for any input store $\store$ and any $\pi_{(\sigma,\store',\callcc\proc(\bar{\clos},\bar{\term}))} \sqsubseteq \pi_{(\emptyset,\store,\prog)}$, $\#\pi_{(\sigma,\store',\callcc\proc(\bar{\clos},\bar{\term}))} \leq Q (m_{\store}^{\prog,\proc}).$
\end{definition}

A procedure with Polynomial Step Count terminates in time polynomial in the size of the initial store and the maximal size of an oracle answer.

\begin{theorem}\label{thm:pol}
Each procedure $\proc$ of a program in $\safe \cap \ASN$ has Polynomial Step Count.
\end{theorem}

\begin{proof}
Let $\prog\in \safe\cap\ASN$.
Now consider a procedure $\proc$ of $\prog$.
Consider an outermost while loop $\rho \rightslice \pbl  \while (\e) \{\cmd\}:\sla$ of a procedure $\proc$ in $\prog$. For any subexpression $\e' \in \e$, by Lemma~\ref{lem:ss2}, either the level of $\e'$ is greater than the level $\sla$ of $\e$ or $\e'$ is declassified in $\e$. Hence $\e$ can be written as follows $e=C[\bar{\declass(\e_1,\e_2)}]$, for some context $C$ with no declassification (note: $C$ is the straightforward extension of contexts to multiple hole contexts). In particular, this implies that any expression $\e'$ in $C$ has level greater than $\sla$.
By Lemma~\ref{lem:stab2}, for any expression $\e'$ in $C$ and any store $\store'$ (obtained from the input store $\store$) if $(\sigma,\store',\phi,\e') \toexp \w$ then $\w \in \Span$.
By the semantics of $\declass$ and by Corollary~\ref{coro:bound2}, it also holds that each $\declass(\e_1,\e_2)$ is evaluated to a unary value in $\Span$, for each store obtained from the input store $\store$.
Consequently, the number of distinct values that $\e$ can take is then bounded by $(\#\Span)^c$, for some constant $c$ bounded by the size of the program, which is polynomial in $\size{\store}$.
Indeed, if the loop is to be taken more than $(\#\Span)^c$ times, it means that the same configuration for variables of level at least $\sla$ in $C$ and $\declass(\e_1,\e_2)$ is visited twice, contradicting the II-aperiodicity hypothesis. Consequently, this while loop can be unfolded polynomially many times. By Lemma~\ref{lem:ci2}, the values stored in statements of level strictly smaller than $\sla$ can increase at most polynomially (i.e., polynomially many times by a constant) in $\size{m_{\store}^{\prog}}$. The same reasoning can be performed on nested loops of level smaller (non-strictly) than $\sla$. Hence the outermost loop has Polynomial Step Count, where the polynomial is a composition of finitely many polynomials, whose number is bounded by the number of nested loops.
Finally, by an induction on the number of sequential outermost loops in the procedure $\proc$, we conclude that $\proc$ has Polynomial Step Count.
\qedhere
\end{proof}

A direct corollary is that each procedure call of a program in $\safe \cap \ASN$ can be simulated by an OTM in $\opt$.

\subsubsection*{Finite Length Revision.}\label{ss:flr}
In this section, we show that aperiodic, terminating, and safe programs can only perform a constant number of oracle calls whose answer is  of increasing size with respect to previous calls.

Given a sequence $(u_n)$, define the length revision of $(u_n)$ as $lr((u_n)) \triangleq \#\{i \ | \ \size{u_i} >\max_{j < i}(\size{u_j})\}.$
Given a program $\prog$, an input store $\store\triangleq \store_\emptyset[\bar{\x} \leftarrow \bar{\w},\bar{\X} \leftarrow \bar{f}]$, a procedure $\proc$ of $\prog$, and an evaluation tree $\pi_{(\sigma,\store',\callcc\proc(\bar{\clos},\bar{\term}))} \sqsubseteq \pi_{(\emptyset,\store,\prog)}$, let $(l^{\pi_{(\sigma,\store',\callcc\proc(\bar{\clos},\bar{\term}))}}_n)$ be the sequence of output values $\w$ in (Orc) rules in $\pi_{(\sigma,\store',\callcc\proc(\bar{\clos},\bar{\term}))}$ except those appearing in a subtree of root indexed by Rules (Brk$_\top$) or (Brk$_\bot$), and that are obtained by a left-to-right depth-first traversal of the evaluation tree $\pi_{(\sigma,\store',\callcc\proc(\bar{\clos},\bar{\term}))}$.

Note the left-to-right depth-first traversal in a derivation exactly corresponds to the order of a sequential execution of a statement.
Hence $(l^{\pi_{(\sigma,\store',\callcc\proc(\bar{\clos},\bar{\term}))}}_n)$ corresponds to the sequence of calls to oracles in $\store$ that do not occur in the guard of a break statement. Oracle answers in break statements do not need to be taken into account for the result to hold as it is only necessary to read a fixed size prefix of the oracle answer to decide whether to break.

\begin{definition}
A procedure $\proc$ of the program $\prog$ has \emph{Finite Length Revision} if there is a constant $n\in\mathbb{N}$ such that for each input store $\store$ and any $\pi_{(\sigma,\store',\callcc\proc(\bar{\clos},\bar{\term}))} \sqsubseteq \pi_{(\emptyset,\store,\prog)}$, $lr((l^{\pi_{(\sigma,\store',\callcc \proc(\bar{\clos},\bar{\term}))}}_n)) \leq n$.
\end{definition}

\begin{theorem}\label{thm:flr}
Each procedure of a program in $\safe $ has  Finite Length Revision.
\end{theorem}

\begin{proof}
By definition of guarded programs, inside a loop, any call to an oracle $\X(\bar{\e})$ (not appearing in the guard of a break statement) is preceded by a $\brk(\size{\X(\bar{\e})} >\size{\X(\bar{\x})})$ statement. Hence the loop execution is not broken if and only if the size of the value of $\X(\e)$ does not exceed the size of the value of $\X(\bar{\x})$. By typing rule (OBK), the level of $\bar{\x}$ is strictly greater than the outermost level of the loop. Hence, by Lemma~\ref{lem:confinement}, $\bar{\x}$ cannot be updated in the loop. Consequently, the size of oracle calls cannot increase in a loop. The only oracle calls that may increase occur outside loops and their number is constant in the size of the program.\qedhere
\end{proof}

As a direct corollary, each procedure call of a program in $\safe $ can be simulated by an OTM that has $\flr$.

\subsubsection*{Soundness and Completeness.}\label{s:sound}
In this section, we show a soundness and completeness for $\BFF$, using the characterization of Theorem~\ref{thm:KS}, based on $\spt$. 

As a direct consequence of Theorem~\ref{thm:pol} and Theorem~\ref{thm:flr}, we obtain the following result.
\begin{proposition}\label{prop:spt}
Each procedure call of a program in $\safe  \cap \ASN$ can be simulated by an OTM in $\spt$.
\end{proposition}
Given a procedure call $\callcc\proc(\bar{\clos},\bar{\term})$, we will denote by $\sem{\callcc\proc(\bar{\clos},\bar{\term})}$ the functional in $\spt$ computed by the OTM simulating this call.
Now we can deduce from Proposition~\ref{prop:spt} that programs in $\safe  \cap \ASN$ are sound for $\BFF$.


\begin{theorem}[Soundness]\label{thm:soundness}
$\sem{\safe  \cap \ASN} \subseteq \BFF$.
\end{theorem}

\begin{proof}
We define a map $[-]^\dagger$ from programs in $\safe  \cap \ASN$ to functionals in $\lambda(\spt)_2$ as follows:
\begin{align*}
[\va]^\dagger &= \va \\
[\callcc\proc(\bar{\clos},\bar{\term^0})]^\dagger &= \sem{\callcc\proc(\bar{\clos},\bar{\term})} \\
[\letin{{\procn}} \prog]^\dagger &= [\prog]^\dagger \\ 
[\boite{[\va]}{\prog}]^\dagger &= \lambda \va.[\prog]^\dagger
\end{align*}

By Proposition~\ref{prop:spt}, it holds that $[\prog]^\dagger \in \lambda(\spt)_2$. 
Moreover, by construction $\sem{\prog}=[\prog]^\dagger$. Hence, by Theorem~\ref{thm:KS}, $\sem{\proc} \in \BFF$. \qedhere
\end{proof}

\begin{lemma}\label{lem:lambdastab}
$$\lambda(\sem{\safe  \cap \ASN})_{2} \subseteq \sem{\safe  \cap \ASN}.$$
\end{lemma}
\begin{proof}
For a given type $\tau$, let $[\tau]$ be the simple type obtained as follows $[\W] = \TW$ and $[\tau \to \tau'] = [\tau ] \to [\tau']$. 

Let us consider a term of $\lambda(\sem{\safe  \cap \ASN})_{2}$. It has type $\overline{(\W \to \W)} \to \bar{\W} \to \W$.
It can be written as $\lambda \bar{X}.\lambda \bar{x}.t$, with $t$ an order-$0$ term.
This term $t$ can be normalized in order to contain no occurrence of an abstraction $\lambda$ (as in \cite[Theorem 5.12]{CU93}).
This means that in subterms of the form $r\ s$, the function $r$ is of order $1$, hence has type $\tau\to\W$.
We declare procedures to apply a closure to its arguments:

\begin{minipage}{0.42\textwidth}
\begin{lstlisting}[numbers=none]
$\niceinstr{declare}$app(X, x) {
    var z;
    z$^\levela\asg$X(x)
    return z
}
\end{lstlisting}
\end{minipage}

We define a straightforward transformation $[\cdot]$ from functions in $\lambda(\sem{\safe  \cap \ASN})_2$ to normalized terms as follows:
\begin{align*}
[x^\tau] &\triangleq \va : [\tau]\\
[\lambda x^\tau.t^{\tau'}] &\triangleq \abs{[x^\tau]}.[t^{\tau'}] : [\tau \to \tau']\\
[t^{\tau \to \W}\ s^{\tau}] &\triangleq \callcc \mathtt{app}([t^{\tau \to \W}], [s^{\tau}]) :\TW\\
[F^{\overline{(\W \to \W)} \to \bar{\W} \to \W}]&\triangleq \term, \text{with } F \in \sem{\safe  \cap \ASN},
\end{align*}
for $\term$ such that $\sem{\boite{[\bar{\Y},\bar{\y}]}{\letin{\bar{\procn}} \term}} = F$.
The existence of $\term$ is ensured by the definition of $\sem{\safe  \cap \ASN}$.
Given $f \in \lambda(\sem{\safe  \cap \ASN})_2$, let $[f] = \lambda\bar{X}.\lambda \bar{x}.[t]$ be a term obtained by applying the above transformation and let $p(f)$ be the list of procedure declarations obtained for each $F$ in the above transformation additioned with the procedure declaration for \verb:app:.
Notice that the existence of $p(f)$ is ensured by definition.
The program $$\prog \triangleq \boite{[\bar{\X},\bar{\x}]}{\letin{p(f)} [t]}$$ is such that $\sem{\prog} =\lambda \bar{X}.\lambda \bar{x}.t  = f$.
$\prog$ is trivially in $\safe  \cap \ASN$ as the procedures are left unchanged, their semantics is already encompassed by the semantics of the original program, and the simply-typed lambda closure is terminating.
\qedhere
\end{proof}

\begin{lemma}\label{lem:ex:safe}
Program $\mathtt{I}$ of Listing~\ref{fig:bruce} is a safe program.
\end{lemma}
\begin{proof}
The innermost/outermost level of $\e$ (resp: $\cmd$) will be the level of the guard of the innermost/outermost while loop containing $\e$. If $\e$ is not in a while then the outermost level is equal to $\levela$ and the value of the innermost level does not matter.  In Listing~\ref{ex:K}, procedure \verb!K! body can be typed as follows:\\
 
\begin{minipage}{0.42\textwidth}
\begin{lstlisting}[firstnumber=22]
$\while$(s$^{\levelb}$>0)$^{\levelb}${
  $\brk$(|X(i)$^{\infty}$|>|X(r$^{\levelc}$)|): $\levelb\sep$
  i$^{\levela}\asg\truncate$(X(i)$^{\infty}$,p$^{\levelb}$)$^{\levela}: \levela\sep$
  s$^{\levelb}\asg$(s$^{\levelb}$-1)$^{\levelb}$ : $\levelb$
}
 \end{lstlisting}
 \end{minipage}
 
The level of \verb!s! at line 22 is enforced to be at least $\levelb$ (and cannot be $\infty$), by Rule (WI). Hence in the while loop body the innermost and outermost levels are equal to $\levelb$. 
Consequently, the level of \verb!r! at line 23 is enforced to be $\levelc$, as it needs to be strictly greater than the outermost level, by Rule (OBK). 
The level of \verb!p! at line 24 is enforced to be equal to the outermost level $\levelb$, by safety on $\truncate$. 
The level of the whole expression $\truncate$\verb!(X(i),p)! is $\levela$. 
Indeed, it has to be smaller than the innermost level $\levelb$. 
Hence at line 24, the level of \verb!i! is enforced to be $\levela$ by Rule (ASG) and the statement can be given the level $\levela$. 
Finally, the full loop body can be typed using Rules (SEQ) and (SUB). 
Notice that the remaining lines of Listing~\ref{ex:K} type as there are no constraints on assignments outside loops.

To conclude, we have typed procedure \verb!K! using the rules of Figure~\ref{sfig:tsprog} with respect to a variable typing environment $\typenv$ defined by $\typenv(\verb!i!)\triangleq \levela$, $\typenv(\verb!s!)=\typenv(\verb!p!)\triangleq \levelb$, $\typenv(\verb!r!)\triangleq \levelc$ (the level of \verb!j! does not matter) and with respect to the operator typing environment $\typop$ satisfying:
$
\levelb \to \levelb \in \typop(-1)(\levelb,\levelb) \cap \typop(>0)(\levelb,\levelb).
$
This constraint holds as $-1$ and $>0$ are neutral. We conclude by observing that $\typenv,\typop \vdash_\levela^\levela \body(\verb!K!) : \levelb$ can be derived.
Now consider the procedure \verb!J! of Listing~\ref{ex:J}.\\

\begin{minipage}{0.42\textwidth}
 \begin{lstlisting}[firstnumber=8]
$\while$(n$^{\levelb}$>0)$^{\levelb}${
  $\brk$(|$\Y$(l$^{\levela}$,n$^{\levelb}$)|>|$\Y$(l0$^{\levelc}$,n0$^{\levelc}$)|) : $\levelb$$\sep$
  t$^{\levela}\asg\truncate$(Y(l,n),b$^{\levelb}$)$^{\levela}$ : $\levela\sep$
  l$^{\levela}\asg\gauche(\droite$(t$^{\levela}$))$^{\levela}$ : $\levela\sep$
  n$^{\levelb}\asg\declass(\droite$($\droite$(t))$^{\levela}$,b$^{\levelb}$)$^{\levelb}$ : $\levelb$
}
 \end{lstlisting}
 \end{minipage}

The level of \verb!n! at line 8 is enforced to be at least $\levelb$ (and cannot be $\infty$), by Rule (WI).
Hence in the while rule body the innermost and outermost levels are equal to $\levelb$. 
At line 9, the level of \verb:l0: and \verb:n0: is enforced to be at level $\levelc$ to be strictly greater thant the outermost level.
Consequently, the level of \verb!b! at line 10 is enforced to be at least $\levelb$.
The level of \verb!t! at line 10 is enforced to be $\levela$ by Rule (ASG).
	Finally, the statement at line 12 needs a $\declass$ as it declassifies data from level $\levela$ (\verb:v:) to level $\levelb$ (\verb:n:).
The body of the loop can be typed using Rules (SEQ) and (SUB).
To conclude, we have typed procedure \verb!J! using the rules of Figure~\ref{sfig:tsprog} with respect to a variable typing environment $\typenv'$ defined by $\typenv'(\verb!n!)=\typenv'(\verb!b!) \triangleq \levelb$ and $\typenv'(\verb!t!)=\typenv'(\verb!l!)=\typenv'(\verb!z!) \triangleq \levela$ (the level of \verb!m! does not matter) and with respect to the operator typing environment $\typop$ satisfying 
$\levelb \to \levelb \in \typop(=0)(\levelb,\levelb) \cap \typop(>0)(\levelb,\levelb)$ and $\levela \to \levela \in \typop(\gauche)(\levelb,\levelb) \cap \typop(\droite)(\levelb,\levelb) \cap \typop(\pars)(\levelb,\levelb)$.
This constraint holds as all these operators are neutral. We conclude by observing that the judgment $\typenv,\typop \vdash_\levela^\levela \body(\verb!J!) : \levelb$ can be derived.
Finally, for the program {\tt I} of Listing~\ref{fig:bruce}, we define the simple typing environment $\typenvs\triangleq\{\texttt{Y}:\TW\to\TW, \texttt{a,b,c}: \bar{\TW}\}$ and the procedure typing environment $\typproc$ by $\typproc({\tt K})\triangleq  \langle \typenv , (\levelb,\levela,\levela)\rangle$ and $\typproc({\tt J})\triangleq  \langle \typenv', (\levelb,\levela,\levela)\rangle$.
The judgment  $\emptyset,\typproc,\typop \vdash {\tt I} : (\TW \to \TW) \to \TW^3 \to \TW$ can be derived and, consequently, the program ${\tt I}$ is safe.  
\end{proof}


\begin{restatable}[Completeness]{theorem}{thmcompleteness}\label{thm:completeness}
$\BFF \subseteq \sem{\safe  \cap \ASN}$.
\end{restatable}

\begin{proof}
The bounded iterator $\mathcal{I}$ is computed by program \texttt{I} from Listing~\ref{fig:bruce}.
We have shown that ${\mathcal{I}} \in \sem{\safe \cap \ASN}$ in Example~\ref{ex:sns} and Lemma~\ref{lem:ex:safe}.
\begin{align*}
\BFF & = \lambda({\FP} \cup \{\mathcal{I}\})_2 && (\text{from~\cite{CU93}})\\
&\subseteq \lambda(\sem{\safe  \cap \ASN})_2 &&(\text{using Theorem}~\ref{thm:fpcomplete})\\
&\subseteq \sem{\safe  \cap \ASN} && (\text{by Lemma}~\ref{lem:lambdastab})
\end{align*}
Note that Theorem~\ref{thm:fpcomplete} can be used as those programs are included in the higher order language and safe aperiodic terminating implies belonging to $\safe\cap\ASN$.
\end{proof}

\thmmain*
\begin{proof}
Direct from Theorems~\ref{thm:soundness} and~\ref{thm:completeness}.
\end{proof}



%
\section{Type Inference and Aperiodicity}

\thmtiproc*

\begin{proof}
The proof of this result uses a reduction to 2-SAT with $\mathcal{O}(\size{\cmd}\times \sla^2)$ clauses, with $\sla$ the number of different levels, $\sla \leq \size{\cmd}$.
 This instance can be solved in time linear in the number of clauses~\cite{EveItaSha76,AspPlaTar79}.

Assume that the number of levels distinct from $\infty$ is fixed, i.e., each level will be in $\{\levela, \levelb, \dots, \sla-1, \infty\}$, for some $\sla$.
Any level can be represented by $\sla+1$ boolean variables $x_{\levela}, x_{\levelb}, \dots, x_{\sla}$ such that $\neg x_{\levela}\wedge \neg x_{\levelb}\wedge\dots\wedge \neg x_{\textbf{i}-1} \wedge x_{\textbf{i}}\wedge\dots\wedge x_{\sla-1}\wedge x_{\sla}$ encodes the level $\textbf{i}$ (and $\sla$ will represent level $\infty$).
The fact that such a set of boolean variables correctly encodes a level can be ensured using $\bigwedge_{\levela\leq j<\sla} \neg x_{j}\vee x_{j+1}$, which accounts for $\sla$ 2-clauses.
We use such boolean variables for each variable in the procedure.
For typing each expression, we use the same construction, but we add another boolean variable to encode the value of $l$.
The levels $\sla_{in}$ and $\sla_{out}$ under which each statement and expression is to be typed will also be encoded similarly.

So for each variable, expression and statement, we use a number of clauses that is linear in $\sla$.

Then each application of a derivation rule in the typing tree will create constraints on the levels that can be either equalities, inequalities or finiteness of levels.
Those can be encoded by a linear in $\sla$ number of 2-clauses.

For a given $\sla$, we obtain a 2-SAT instance with $O(\size{\cmd}\times \sla^2)$ clauses.
The typability of $\cmd$ implies that the number of levels is bounded by $\size{\cmd}$, hence the type inference can be done in time $O(\size{\cmd}^3)$.
\end{proof}

\thmap*

\begin{proof}
Aperiodicity is a $\Pi^0_1$ property: it can be encoded by a formula meaning for all input, for all numbers of steps $n_1$ and $n_2$ such that $n_1\neq n_2$ and $(\store, \cmd)$ reduces to $(\store'_i, \whar{\e}{\cmd})$ in $n_i$ steps, we have $\store'_1 \not \equiv_{\e} \store'_2$.

Aperiodicity is $\Pi^0_1$-hard: let ${\main(\x)\{\cmd\ \ret\ \z\}}$ be a program with while loops.
Let $\widetilde{\cmd}$ be a rewriting of $\cmd$ where each $\whar{\e_i}{\cmd_i}$ is rewritten as $\x_i := 1; \whar{\e_i\ \texttt{and}\ \x_i > 0}{\x_i := \x_i +1; \cmd_i}$ with a fresh variable $\x_i$.
The program $\texttt{p}(\x)\{\whar{1}{\widetilde{\cmd}}\ \ret\ \z\}$ is aperiodic if and only if $\main$ does not halt.
On the whole, if $\main$ does not halt, then $\texttt{p}$ is aperiodic, indeed because of the $\x_i$, no aperiodicity occurs inside $\widetilde{\cmd}$, and the outer $\while(1)$ will be reached only once; otherwise, if $\main$ halts, $\texttt{p}$ is not aperiodic ($\while(1)$ will be seen more than once).
\end{proof}

\thmtiprog*

\begin{proof}
Let $\prog$ be a program and $\typop$ a safe operator typing environment.
To show that $\prog$ is safe, we need to produce a procedure typing environment $\typproc$ such that $\emptyset, \typproc, \typop \vdash \prog : \overline{\TW \to \TW} \to \bar{\TW} \to \TW$ can be derived.
Let us assume that $\prog =\boite{[\bar{\X},\bar{\x}]}{\letin{\bar{\procn}} \term}$.
After $\ell(\bar{\X},\bar{\x})$ applications of the Rule (BOX) of Figure~\ref{sfig:tsprog}, we encounter the following judgment
$\{\bar{\X} : \overline{\TW \to \TW},\bar{\x} : \bar{\TW}\},\typproc,\typop \vdash  \letin{\bar{\procn}} \term:  \TW.$
Deriving the above judgment consists of deriving:
$\pi \rightslice \ \{\bar{\X} : \overline{\TW \to \TW},\bar{\x} : \bar{\TW}\},\typproc,\typop \vdash  \term:  \TW$ and
for each $1 \leq i \leq \ell(\bar{\procn})$ judgments of the shape:
$\pi_i \rightslice \typproc_1^i,\Delta \vdash \body(\procn_i) : \typproc_2^i$
where $\langle \typproc_1^i, \typproc_2^i\rangle \triangleq \typproc(\procn_i)$,
after $\ell(\bar{\procn})$ applications of the Rule (DEF) of Figure~\ref{sfig:tsprog}.
By Theorem~\ref{thm:tiproc}, for each derivation $\pi_i$, type inference can be done in time $\mathcal{O}(\size{\procn_i}^3)$.
Hence, all these judgments can be inferred in time $\mathcal{O}(\size{\prog}^3)$.

It remains to study the type inference problem for the derivation $\pi$.
By looking at Figure~\ref{sfig:tsprog}, the typing discipline for terms follows a standard simply-typed discipline augmented with constants (the procedure calls).
It is well-known that type inference in the simply-typed lambda-calculus is a $\Ptime$-complete problem as any instance of the Circuit Value Problem (CVP) can be encoded in the former~\cite{M04}.
The typing rules presented in the lower part of Figure~\ref{sfig:tsprog} are just a restriction of simple typing to second order.
\end{proof}

\end{document}